\documentclass[fleqn,usenatbib]{mnras}
\usepackage{fix-cm}
\usepackage{graphicx}
\usepackage{subcaption}
\usepackage{float}
\usepackage{gensymb}
\usepackage{newtxtext,newtxmath}
\usepackage{esint}
\usepackage{mathtools}

\usepackage[T1]{fontenc}
\newcommand{\HI}{H\,\textsc{i}}
\newcommand{\hi}{H\textsc{i}}
\defcitealias{Meer21cm}{Chen and Cunnington et~al.}
\newcommand{\citepaliasyear}[1]{(\citetalias{#1}~\citeyear{#1})}   

\DeclareRobustCommand{\VAN}[3]{#2}
\let\VANthebibliography\thebibliography
\def\thebibliography{\DeclareRobustCommand{\VAN}[3]{##3}\VANthebibliography}

\usepackage{graphicx}	
\usepackage{amsmath}	

\newcommand{\secref}[1]{\hyperref[#1]{Section~\ref*{#1}}}
\newcommand{\appref}[1]{\hyperref[#1]{Appendix~\ref*{#1}}}

\makeatletter
\patchcmd{\ps@titlepage}
  {\def\@oddfoot{\small\copyright\ \@pubyear\ The Authors\hfil}}
  {\def\@oddfoot{%
     \small\copyright\ \@pubyear\ The Authors\hfill
     \begin{minipage}[b]{\columnwidth}%
       \raggedright\footnotesize
       \hypertarget{corautbz}{\textcolor{\@linkcolor}{$^{\star}$}}\hspace{0.35em}E-mail: zhaoby@nao.cas.cn\\
     \end{minipage}}}
  {}{\errmessage{Could not place corresponding-author emails}}
\makeatother

\title[Oscillatory systematics in MeerKLASS]{Mitigating oscillatory systematics in MeerKLASS L-band \HI\ intensity maps}

\author[B. Zhao et al.]{
Boyan Zhao$^{1,2,3}$\textsuperscript{\hyperlink{corautbz}{$\star$}},
Eloy de Lera Acedo$^{2,4}$,
Zhaoting Chen$^{5,6}$,
John Cumner$^{2,4}$,
Vijay Mahatma$^{2,4}$,
\newauthor\ 
Marta Spinelli$^{6,7}$,
Dirk I. L. de Villiers$^{8}$,
Matilde Barberi-Squarotti$^{9,10,11,12}$,
Phil Bull$^{13,7}$,
\newauthor\ 
Stefano Camera$^{12,14,15,6}$,
Isabella Paola Carucci$^{16,17}$,
Xuelei Chen$^{1,3,18}$,
Steven Cunnington$^{19}$,
\newauthor\ 
Furen Deng$^{1,3}$,
Brandon N. Engelbrecht$^{7}$,
Jos\'e Fonseca$^{20,21,7}$,
Keith Grainge$^{13}$,
Wenkai Hu$^{1,3}$,
\newauthor\ 
Melis O. Irfan$^{22,7}$,
Yichao Li$^{23}$,
Mario G. Santos$^{7}$,
Jingying Wang$^{24,7}$,
Yougang Wang$^{1,3,18}$,
and Laura Wolz$^{13}$
\\
$^{1}$State Key Laboratory of Radio Astronomy and Technology, National Astronomical Observatories, CAS, \\ \ \ A20 Datun Road, Chaoyang District, Beijing, 100101, P. R. China;\\
$^{2}$Cavendish Laboratory, University of Cambridge, Cambridge, CB3 0HE, UK\\
$^{3}$School of Astronomy and Space Science, University of Chinese Academy of Sciences, Beijing 100049, China\\
$^{4}$Kavli Institute for Cosmology in Cambridge, University of Cambridge, Cambridge, CB3 0HE, UK\\
$^{5}$Institute for Astronomy, The University of Edinburgh, Royal Observatory, Edinburgh EH9 3HJ, UK\\
$^{6}$Observatoire de la Côte d’Azur, Laboratoire Lagrange, Bd de l’Observatoire, CS 34229, 06304 Nice cedex 4, France\\
$^{7}$Department of Physics \& Astronomy, University of the Western
Cape, Robert Sobukwe Road, Cape Town 7535, South Africa\\
$^{8}$Department of Electrical and Electronic Engineering, Stellenbosch University, Stellenbosch, South Africa\\
$^{9}$Dipartimento di Fisica, Universit\`a degli Studi di Milano, via G.\ Celoria 16, 20133 Milano, Italy\\
$^{10}$INFN – Istituto Nazionale di Fisica Nucleare, Sezione di Milano, via G.\ Celoria 16, 20133 Milano, Italy\\
$^{11}$INAF -- Istituto Nazionale di Astrofisica, Osservatorio Astrofisico di Brera-Merate, via Brera 28, 20121 Milano, Italy\\
$^{12}$Dipartimento di Fisica, Universit\`a degli Studi di Torino, via P.\ Giuria 1, 10125 Torino, Italy\\
$^{13}$Jodrell Bank Centre for Astrophysics, Department of Physics \& Astronomy, The University of Manchester, Manchester M13 9PL, UK\\
$^{14}$INFN – Istituto Nazionale di Fisica Nucleare, Sezione di Torino, via P.\ Giuria 1, 10125 Torino, Italy\\
$^{15}$INAF – Istituto Nazionale di Astrofisica, Osservatorio Astrofisico di Torino, 10025 Pino
Torinese, Italy\\
$^{16}$INAF -- Istituto Nazionale di Astrofisica, Osservatorio Astronomico di Trieste, Via G.B.\ Tiepolo 11, 34131 Trieste, Italy\\
$^{17}$IFPU -- Institute for Fundamental Physics of the Universe, Via Beirut 2, 34151 Trieste, Italy\\
$^{18}$Key Laboratory of Cosmology and Astrophysics (Liaoning) \& College of Sciences, Northeastern University, Shenyang 110819, China\\
$^{19}$Institute of Cosmology \& Gravitation, University of Portsmouth, Dennis Sciama Building, Portsmouth, PO1 3FX, UK\\
$^{20}$Instituto de Astrof\'isica e Ci\^encias do Espa\c{c}o, Universidade do Porto CAUP, 4150-762 Porto, Portugal\\
$^{21}$Departamento de F\'isica e Astronomia, Faculdade de Ci\^{e}ncias, Universidade do Porto, Rua do Campo Alegre 687, 4169-007 Porto, Portugal\\
$^{22}$Institute of Astrophysics, University of Cambridge, Madingley Road, CB3 0HA, UK\\
$^{23}$Department of Physics, College of Sciences, Northeastern University, Wenhua Road, Shenyang 11089, China\\
$^{24}$State Key Laboratory of Radio Astronomy and Technology, Shanghai Astronomical Observatory, CAS, 80 Nandan Road, Shanghai 200030, China\\
}

\date{Accepted XXX. Received YYY; in original form ZZZ}

\pubyear{\the\year{}}

\begin{document}
\label{firstpage}
\pagerange{\pageref{firstpage}--\pageref{lastpage}}
\maketitle
\begin{abstract}
\HI\ intensity mapping is a promising probe of the large-scale structure of the Universe, but is limited by foregrounds and instrumental systematics.
For the MeerKAT Large Area Synoptic Survey (MeerKLASS) L-band data, it is found that emission line stacking measurement reveals a type of systematics related to the instrument beam \citep{Chen2025}.
Using simulations with a full-sky electromagnetic beam model, we show that the chromaticity and sidelobes of the MeerKAT beam modulate the frequency-smooth foreground emission and introduce oscillating structures that persist after foreground removal.
The beam rotates with respect to the sky between two types of scans, and combining datasets with different beam orientations before foreground cleaning makes the residuals stronger.
In MeerKLASS L-band data, processing these two types of scans separately significantly suppresses the oscillating residuals while preserving the stacked \HI\ signal.
The ratio of the \HI\ signal amplitude to the systematic fluctuation amplitude increases by $81\%$.
We find an oscillating structure in PCA mode 11 in our previous data processing, but simply removing this mode does not improve the stacked spectrum due to the signal loss.
Our strategy can help mitigate spectral ripples that are induced by the beam in future MeerKAT and SKA-Mid surveys.
\end{abstract}

\begin{keywords}
cosmology: observations --
large-scale structure of Universe --
radio lines: general --
methods: observational --
methods: data analysis
\end{keywords}



\section{Introduction}

The large-scale distribution of matter in the Universe, particularly that of dark matter, is one of the primary targets of modern cosmology, as it provides key constraints on cosmological models and fundamental physics.
Although dark matter cannot be detected directly through electromagnetic radiation, its distribution can be inferred from observable tracers.
Among these, the redshifted 21-cm emission from neutral hydrogen (\HI) is particularly promising because most \HI\ resides in dark matter haloes in the post-reionization Universe ($z\lesssim6$; \citealt{Villaescusa-Navarro2018}).
Therefore, the measurement of \HI\ distribution probes the distribution of dark matter and helps us test the cosmological model \citep{Bull2015}.
Since cosmological information is encoded in the large-scale fluctuations of the integrated \HI\ emission, individual \HI-bearing galaxies do not need to be resolved.
Instead, the unresolved \HI\ emission from many galaxies can be mapped over coarsely sampled sky areas, allowing efficient surveys of large cosmological volumes with relatively small-aperture telescopes.
This technique trades angular resolution for survey speed and is known as \HI\ intensity mapping (e.g.\ \citealt{Madau1997,Bharadwaj2001,Chang2008,Wyithe2008}).

One of the major challenges in \HI\ intensity mapping is the removal of foreground contamination, which is dominated by Galactic synchrotron and free--free emission, together with their extragalactic counterparts.
The foreground remains spectrally smooth and is predominantly described by a power law, but its amplitude varies at the Kelvin level across different lines of sight, as shown in \citet{MeerKLASS2025}.
At $z \lesssim 6$, the foreground emission is at least four orders of magnitude brighter than the expected cosmological \HI\ signal \citep{Santos2015, Wolz2026}.
The cosmological \HI\ signal exhibits more rapid frequency fluctuations due to the different frequencies corresponding to the 21\,cm line at different redshifts.
A variety of blind component separation methods have therefore been developed \citep{Spinelli2022}, such as Principal Component Analysis (PCA) and its multi-scale extension \citep{Carucci2025}; Independent Component Analysis (ICA; \citealt{Chapman2012}); Generalized Morphological Component Analysis (GMCA; \citealt{Carucci2020}); and the Generalized Needlet Internal Linear Combination (GNILC; \citealt{2011MNRAS.418..467R,Olivari2016}) method.
A comprehensive overview and comparison of these foreground-cleaning methods is provided in \citet{Spinelli2026}.

Over the past decade, \HI\ intensity mapping observations have been carried out with various radio telescopes, including the Green Bank Telescope \citep{Chang2010,Masui2013,Switzer2013,Wolz2022}, the Canadian Hydrogen Intensity Mapping Experiment \citep{CHIME2022,CHIME2025}, the Five-hundred-meter Aperture Spherical Telescope \citep{Li2023}, and the Tianlai array \citep{Zuo2021}.
Another important facility for \HI\ intensity mapping is MeerKAT, a radio interferometer located in South Africa, consisting of 64 dishes, each 13.5 m in diameter, and serving as a precursor to the Square Kilometre Array (SKA).
MeerKAT can perform \HI\ intensity mapping in both interferometric and single-dish modes, allowing it to probe a wide range of spatial scales.
In interferometric mode, MeerKAT has been used to measure the \HI\ power spectrum on relatively small scales \citep{Paul2023,Mazumder2025,Townsend2026}.
In single-dish mode, MeerKAT has conducted several \HI\ intensity mapping campaigns, including the MeerKLASS L-band pilot survey \citep{Wang2021,Cunnington2023,Carucci2025} and the subsequent L-band deep-field observations \citep{MeerKLASS2025, MeerKLASS2026}.
In this work, we focus on the latter.
The single-dish observations retain large-scale modes inaccessible to the interferometric measurements, while the long integration time of approximately 62 hours per dish provides improved sensitivity compared with previous MeerKLASS single-dish observations \citep{MeerKLASS2025, MeerKLASS2026}.
However, the performance of foreground-cleaning methods is often degraded by instrumental systematics.
One of the most important is the telescope beam, which exhibits frequency-dependent oscillations and complex sidelobe structures \citep{Matshawule2021}.
These beam effects introduce additional spectral structure into the measured data which cannot be removed by PCA effectively \citep{Spinelli2022,Spinelli2026}.

The quality of foreground removal directly affects all subsequent scientific analyses.
Common applications of \HI\ intensity maps include measurements of the auto-correlation power spectrum (e.g. \citealt{Bharadwaj2001}), the cross-correlation power spectrum with optical galaxies (e.g. \citealt{Chang2010,Masui2013,Switzer2013,Wolz2016,Wolz2017}), and the stacked \HI\ signal at the positions of optically identified galaxies (e.g. \citealt{Zwaan2001,Lah2007,Fabello2011,Delhaize2013,Tramonte2020}).

In \citet{MeerKLASS2025}, the measured auto-power spectrum of the MeerKLASS L-band deep field was found to contain substantially more power than expected.
This excess, which is later mitigated by performing internal cross-correlation between different \HI\ datasets \citep{MeerKLASS2026}, indicates that residual systematics may still be present in the data after foreground cleaning.
However, a power spectrum is a statistical description of the fluctuations in the data and does not directly show the spatial or spectral structures responsible for the excess power.
Stacking provides a complementary way to investigate these structures.
Unlike the power spectrum, the stacking results can be directly examined as angular images and frequency spectra.
This allows us to see how residual systematics vary across the sky and with frequency, providing a more direct way to investigate their properties and possible origins.
Understanding and mitigating these systematics through stacking can therefore benefit not only the stacking measurement itself, but also power-spectrum analyses performed on the same data.

A first step in this direction was made by \citet{Chen2025}.
They found a strong oscillating structure in the stacked \HI\ signal, together with correlations between neighbouring frequency channels. 
This is shown to be caused by the chromatic beam response, but the specific mechanism remains unclear.
Meanwhile \citet{Cerfonteyn2024} have studied receiving sensitivity ripple in electrically small offset Gregorian reflector antennas such as MeerKAT.
Building on these works, we aim to elucidate how the beam gives rise to oscillations in the stacked spectrum and to explore methods for mitigating these effects.
The impact of beam effects on foreground cleaning in MeerKAT \HI\ intensity mapping has been investigated by \citet{Matshawule2021}, who simplified the beam as circularly symmetric.
In this paper, we use a new MeerKAT beam model, which provides full-sky coverage and more realistic sidelobes while retaining the asymmetry.
The updated beam model, combined with the stacking analysis pipeline presented in \citet{Chen2025}, allows us to simulate the origin of the systematics and explore an optimal mitigation strategy.

The remainder of this paper is organized as follows.
In \secref{section:Stacking procedure with MeerKLASS L-band data}, we briefly introduce the data and the stacking procedure.
In \secref{section:Simulation}, we describe the beam model, the sky model, and the convolution approach in simulation.
\secref{section:Properties of the systematics} presents the properties of the systematics in the data, in relation to the instrument beam.
Simulation results are presented in \secref{section:Simulation Results}, followed by the analysis of the real data in \secref{section:Improved stacked spectrum with L-band data}.
The conclusion is in \secref{section:Conclusion}.

\begin{figure*}
    \centering
    \includegraphics[width=1\linewidth]{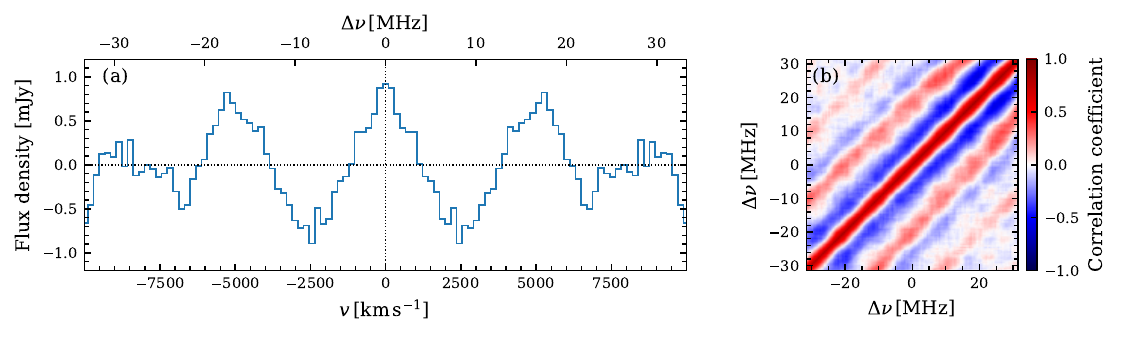}
    \caption{The symmetrised stacked spectrum (left) and correlation coefficient matrix of the reference spectra (right), obtained from the final cube after removing 10 PCA modes following the analysis of \citet{Chen2025}.
    The stacked spectrum exhibits an oscillatory structure, while the correlation matrix shows a ripple pattern, indicating the presence of residual systematics in the final cube.
    These results correspond to those shown in Fig.~16 of \citet{Chen2025}.}
    \label{fig:old_results}
\end{figure*}

\section{Stacking procedure with MeerKLASS L-band data}
\label{section:Stacking procedure with MeerKLASS L-band data}

\subsection{Data}
\HI\ stacking combines the \HI\ signals from many known galaxies after aligning them in position and frequency.
For the \HI\ data, we use the MeerKLASS L-band Deep-field observations taken in 2021 \citep{MeerKLASS2025}.
For galaxy data, we use a spectroscopic galaxy catalogue from the Galaxy And Mass Assembly (GAMA) survey \citep{Driver2009,Driver2011,Liske2015}.

The MeerKLASS L-band deep-field data cover a sky field of $236\,\mathrm{deg}^2$, spanning R.A. $(330^\circ, 360^\circ)$ and Dec. $(-36^\circ, -25^\circ)$.
To obtain a good coverage and high signal-to-noise ratio, the target field was scanned repeatedly.
Some scans were taken when the target field was rising (\textit{rising scans}), while others were taken when it was setting (\textit{setting scans}).
The target field turns ``upside down'' relative to the telescope when it changes from rising to setting.
A map of routes for these two scans can be found in Fig.~3 of \citet{Wang2021}.

The frequency channel width is $\delta\nu \approx208.98\,\mathrm{kHz}$.
Before radio-frequency interference (RFI) flagging, the total integration time per dish is 62 hours.
After flagging, the clean sub-band used in this work is $971.15\,\mathrm{MHz} < \nu < 1023.61\,\mathrm{MHz}$,
corresponding to $0.39 \lesssim z \lesssim 0.46$, with an effective redshift of $z=0.424$ at the centre.
The total integration time of each voxel (i.e. a pixel at a given frequency) is recorded in the hit-count file, where each count corresponds to an integration time of $\delta t=2\,\mathrm{s}$.

The specific area of the GAMA survey used in this work is the 23~hr (G23) field \citep{Driver2022}, which covers $339^\circ < \mathrm{R.A.} < 351^\circ$ and $-35^\circ < \mathrm{Dec.} < -30^\circ$.
In total, $N=2269$ galaxies are selected, with a magnitude limit of $i < 19.2$.
The GAMA galaxy positions as well as the hit-count map can be found in Fig.~1 of \citet{Chen2025}.

\subsection{Stacking procedure}
\label{section:Stacking procedure}
The analysis is performed using the \textsc{Meer21cm} code \citepaliasyear{Meer21cm}.
Unless otherwise stated, we follow the same approach as \citet{Chen2025}.
Here, we briefly describe the stacking procedure.

For each GAMA galaxy, its spectroscopic redshift determines the corresponding \HI\ frequency.
Together with its R.A. and Dec., this locates the galaxy in the MeerKLASS data cube.
A three-dimensional box centred on this position is then extracted.
The extracted boxes are aligned at their centres and combined through a weighted stack.
\begin{equation}
I(\Delta\alpha,\Delta\beta,\Delta\nu)
=
\frac{
\sum_i I\left(\alpha_i+\Delta\alpha,\beta_i+\Delta\beta,\nu_i+\Delta\nu\right)w_i
}{
\sum_i w_i
},
\end{equation}
where $I$ is the flux density, $(\alpha_i,\beta_i,\nu_i)$ gives the (R.A, Dec, frequency) of the voxel that contains the $i$th galaxy in the data cube, 
$(\Delta\alpha,\Delta\beta,\Delta\nu)$ gives the offset from the centre of each box, 
and $w_i$ is the corresponding stacking weight.
The equation therefore averages the voxels at the same relative position in all stacked boxes to produce a three-dimensional stacked cube.
Following \citet{Chen2025}, we adopt inverse noise variance weighting so that $w_i = N^i_{hits}$, where $N^i_{hits}$ are the hit counts.
Hereafter, \emph{cube} refers to the three-dimensional $(\alpha, \beta, \nu)$ map of the full survey area, while \emph{box} refers to a smaller three-dimensional sub-cube extracted from the cube.
\emph{Final cube} refers to the data cube produced by averaging over all dishes and all observations conducted on different nights throughout the survey.
\emph{Rising}/\emph{setting cube} refers to the corresponding data cube produced by averaging over all dishes and all observations from the rising/setting scans, respectively.

The stacked box is three-dimensional $(N_\mathrm{pix} \times N_\mathrm{pix} \times N_\mathrm{freq})$.
To obtain a one-dimensional stacked spectrum, we select a circular region centred on the central pixel and sum the frequency spectra of all pixels within this region (see also Fig.~8 of \citealt{Chen2025}).
The stacked spectrum is
\begin{equation}
\bar{I}(\Delta \nu)=\sum_{n,m} I(\Delta \alpha_n, \Delta \beta_m, \Delta\nu), \quad \text{for } \sqrt{\Delta \alpha_n^2+\Delta \beta_m^2} \leq R_\mathrm{stack},
\label{eq:angular_sum}
\end{equation}
where $(n,m)$ are the indices of the pixels in the angular plane.
Throughout this paper, the radius of this circular region is referred to as the \emph{stacking radius}, $R_\mathrm{stack}$.
The stacked spectrum obtained following the analysis of \citet{Chen2025} is presented in \autoref{fig:old_results}(a).
A peak is expected at $v=0$, corresponding to the stacked \HI\ emission signal.
However, two additional peaks of comparable amplitude are present at $|v|\simeq5000~\mathrm{km\,s^{-1}}$, indicating that substantial residual systematics remain in the stacked spectrum.

To perform a null test, \citet{Chen2025} repeated the same procedure using multiple realizations of randomly generated mock galaxy positions.
These positions follow the same clustering statistics as the real galaxy catalogue but are uncorrelated with the \HI\ signal.
The resulting spectra are referred to as the \emph{reference spectra}.
In the absence of residual systematics, the reference spectra are expected to be noise-like, with a mean consistent with zero and no correlation between different frequency channels.
Instead, clear oscillatory structures are present in the reference spectra.
To quantify these frequency correlations, they construct the correlation coefficient matrix from multiple realizations of the reference spectra,
\begin{equation}
C_{\mu\eta}
=
\frac{
\left\langle
\big(I_\mu-\langle I_\mu\rangle\big)
\big(I_\eta-\langle I_\eta\rangle\big)
\right\rangle
}
{\sigma_\mu\sigma_\eta},
\end{equation}
where $I$ denotes the reference spectrum, $\mu$ and $\eta$ are the frequency-channel indices, $\langle\,\cdot\,\rangle$ denotes the average over all realizations, and $\sigma_\mu$ is the standard deviation of channel $\mu$.
Hereafter, we refer to this simply as the \emph{correlation matrix}.
The diagonal elements represent the self-correlation of each frequency channel, while non-zero off-diagonal elements indicate correlations between different channels.
As shown in \autoref{fig:old_results}(b), the correlation matrix exhibits a clear ripple pattern, demonstrating that periodic frequency correlations remain in the data after PCA foreground cleaning.

This behaviour suggests the presence of an oscillating systematic component in the data, which is thought to originate from the telescope beam, as its period of oscillation matches that expected from diffraction between the secondary reflector and the primary dish of MeerKAT \citep{Dirk2013, Asad2021}.

\section{Simulation}
\label{section:Simulation}
To test whether the oscillatory systematics are associated with the telescope beam, one can convolve a sky model with a beam model and examine whether a similar ripple pattern emerges in the simulated correlation matrix.
However, no such pattern was found in the simulations of \citet{Chen2025}, possibly because their beam model was overly simplified.
It is therefore worth revisiting this test with a more realistic beam model that includes chromatic sidelobes.
In our simulation, we focus on the coupling between foreground emission and the beam response.
Accordingly, no \HI\ signal is included in the simulations.
The simulations are intended to be qualitative rather than to accurately reproduce the observed correlation matrices or stacked spectra.

\subsection{Sky model}
The foreground sky is simulated using \texttt{PySM3} \citep{Thorne2017,Zonca2021,Panexp2025}, including Galactic synchrotron, free-free emission, thermal dust and the CMB.
The synchrotron emission is modelled using the \texttt{s5} model, which adopts a power-law spectrum with a spatially varying spectral index based on the reprocessed 408 MHz Haslam map \citep{Remazeilles2015}.
The free-free emission is simulated using the \texttt{f1} model based on the analytic free-free model \citep{Draine2011} adopted by the \textit{Planck} Commander component-separation analysis.
The CMB is simulated using the \texttt{c1} model based on the best-fitting \textit{Planck} 2015 $\Lambda$CDM cosmology \citep{Planck2016}.
The thermal dust emission, which is sub-dominant, is simulated using the \texttt{d1} model, which adopts a single-component modified blackbody spectrum with parameters derived from the \textit{Planck} Commander component-separation analysis \citep{2016A&A...594A..10P}.
Extragalactic point sources are generated from the NVSS catalogue \citep{1998AJ....115.1693C}, assuming a power-law spectrum with a spectral index of $-0.7$ according to \citet{Miley2008}.

\begin{figure}
    \centering
    \includegraphics[width=1\linewidth]{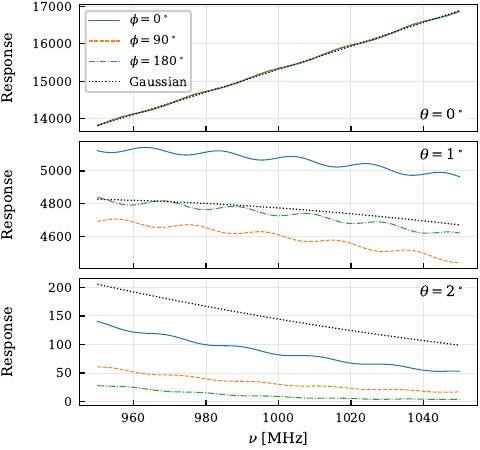}
    \caption{Beam response functions at $\theta = 0^\circ$ (top), $\theta = 1^\circ$ (middle), and $\theta = 2^\circ$ (bottom).
    The black dashed line shows the Gaussian beam model with $\mathrm{FWHM}=1.212\,\lambda/13.5\,\mathrm{m}$, while the solid lines show the Physical Optics (PO) all-sky beam model for three representative azimuthal angles.}
    \label{fig:beam_mainlobe}
\end{figure}

\begin{figure}
    \centering
    \includegraphics[width=1\linewidth]{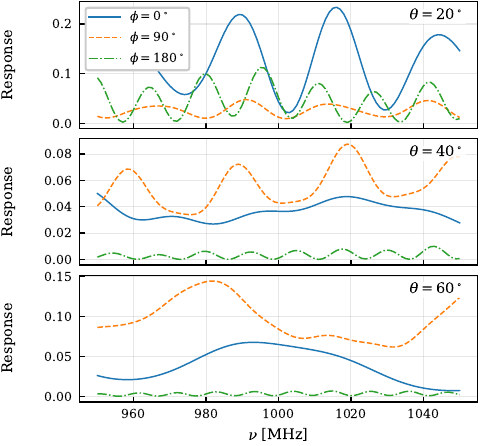}
    \caption{Beam response functions at $\theta = 20^\circ$ (top), $\theta = 40^\circ$ (middle), and $\theta = 60^\circ$ (bottom) for the Physical Optics (PO) all-sky beam model.
    Different directions exhibit distinct beam responses, which introduce additional frequency-dependent structures into the measured spectra.}
    \label{fig:beam_sidelobe}
\end{figure}

\subsection{Beam model}
\label{section:Beam model}
The beam is generated through electromagnetic simulations of a simplified MeerKAT antenna.
For the feed, the orthogonal mode transducer (OMT) is omitted from the model, and only the rotationally symmetric corrugated waveguide horn feed is retained.
A frequency-domain finite-element method (FEM) solution of the feed pattern, assuming fundamental-mode excitation of the antenna waveguide port, is calculated using CST Microwave Studio\footnote{\url{https://www.3ds.com/products/simulia/cst-studio-suite}} and exported as a far-field source.
This source is then used to illuminate a Physical Optics (PO) model of the dual-reflector system in TICRA Tools GRASP\footnote{\url{https://www.ticra.com/software/grasp/}}.
Only the reflecting surfaces are included in the simulation, without the struts or detailed feed structure, and a one-directional fully illuminated PO analysis is performed.
Consequently, multiple-reflection coupling between the reflectors or between the reflectors and the feed structure is not included.
Despite these simplifications, much of the beam chromaticity is captured by the model, since a substantial component of the frequency-dependent structure arises from interference involving fields diffracted by the sub-reflector and main reflector \citep{Dirk2013,Cerfonteyn2023,Cerfonteyn2024}.
Owing to the clear-aperture optics of the offset Gregorian reflector configuration, standing-wave effects between the feed and reflector, which are prominent in prime-focus and symmetric dual-reflector systems, are substantially reduced \citep{Cerfonteyn2023,Cerfonteyn2024}.
Hereafter, we shall refer to this model as the Physical Optics (PO) all-sky beam, or simply the PO beam.

The PO beam model is generated in a spherical coordinate system, where $\theta$ is the polar angle and $\phi$ is the azimuthal angle.
The $\theta = 0^\circ$ direction corresponds to the beam boresight, i.e. the centre of the main lobe.
In \autoref{fig:beam_mainlobe}, we compare the PO beam model with a Gaussian beam model in the vicinity of the beam centre.
In \autoref{fig:beam_sidelobe}, we extend this comparison to the sidelobe region, using $\theta=20^\circ$, $40^\circ$, and $60^\circ$ as representative directions.
The beam response $B(\theta,\phi,\nu)$ is normalized so that
\begin{equation}
    \frac{1}{4\pi}\oiint B(\theta,\phi,\nu)\,\sin\theta\,\mathrm{d}\theta\,\mathrm{d}\phi = 1.
\label{eq:beam_normalization}
\end{equation}
The $\mathrm{FWHM}$ of the Gaussian model is set to $1.212\,\lambda/D$, where $D = 13.5\,\mathrm{m}$ is the diameter of a MeerKAT dish.
We choose the scaling parameter to be 1.212 in this comparison, to make the Gaussian model match the PO model at $\theta = 0^\circ$.
The black dashed line shows the Gaussian beam, while the solid lines show the PO model in three representative directions, where the azimuth $\phi = 180^\circ$ points towards the sub-reflector.

Compared to the Gaussian model, the PO model exhibits a clear oscillatory frequency dependence, with a characteristic period of approximately $20\,\mathrm{MHz}$ near the main lobe.
The oscillation amplitude is about $0.2\%$ at the polar angle $\theta = 0^\circ$ and increases to approximately $0.4\%$ at $\theta = 1^\circ$, roughly corresponding to the edge of the main lobe.
At $\theta = 2^\circ$, the amplitude increases substantially, reaching nearly $5\%$.

The MeerKAT beam is anisotropic in the $\phi$ direction, as demonstrated by \citet{Asad2021} and \citet{Dirk2013}.
This anisotropy is also captured by the PO beam model adopted in this work.
As shown in \autoref{fig:beam_mainlobe} and \autoref{fig:beam_sidelobe}, the beam response varies with $\phi$ even at fixed $\theta$.
For example, at $\theta = 60^\circ$, a prominent oscillation with a period of approximately $10\,\mathrm{MHz}$ is present at $\phi = 180^\circ$.
In contrast, at $\phi = 0^\circ$ and $90^\circ$, this $10\,\mathrm{MHz}$ oscillation disappears, leaving only variations on much broader frequency scales.
In addition, the beam response also varies strongly with $\theta$.

\begin{figure}
    \centering
    \includegraphics[width=\linewidth]{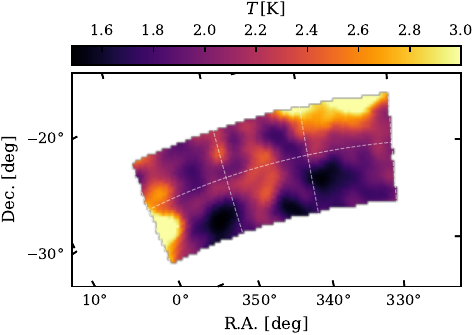}
    \caption{Simulated sky map at $1000\,\mathrm{MHz}$ after convolution with the PO beam model. The input sky consists of synchrotron radiation, free-free emission, thermal dust emission, CMB anisotropies from \texttt{PySM3}, and NVSS point sources. The beam integration is performed with $\theta_{\mathrm{max}} = 90^\circ$.}
    \label{fig:map_sim}
\end{figure}

\begin{figure}
    \centering
    \includegraphics[width=1\linewidth]{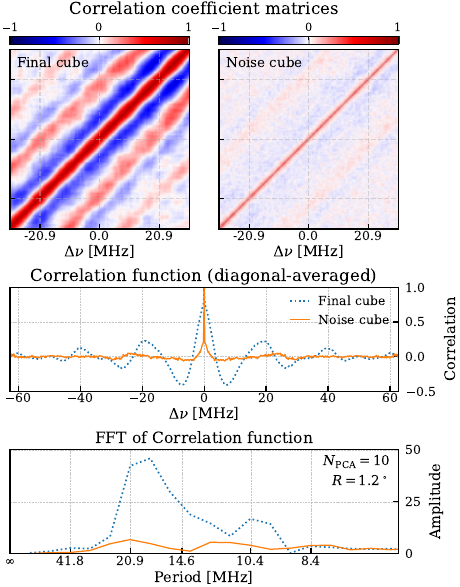}
    \caption{Upper panel: correlation matrices of the data cube (left) and a mock noise cube (right). The first 10 PCA modes are removed from the final cube and the same 10 modes are projected out from the noise cube. The correlation in the noise cube is significantly weaker than that in the final cube. 
    Middle panel: the normalized two-point correlation functions obtained by averaging the matrices along their diagonals.
    Lower panel: the Fourier transforms of the corresponding correlation functions, showing the characteristic period of the oscillating structure.}
    \label{fig:matrix_noise}
\end{figure}

\subsection{Beam convolution}
The sky map is convolved with the PO beam to simulate the sky observed by MeerKAT.
The radiative transfer through the atmosphere and the ground reflections are not included because our aim is to gain insights from simulation to improve the processing of real data, instead of exactly reproducing the observational results.

The PO beam is sampled on a spherical grid as $B(\theta_i,\phi_j)$, whereas the sky temperature is provided as a HEALPix\footnote{\url{http://healpix.sourceforge.net}} map \citep{2005Healpy,2019Healpy}, $T_{\mathrm{pix}}$.
First, we resample the sky temperature onto the beam grid for each pointing direction $\boldsymbol{r}_{\mathrm p}$ through a rotation matrix $\mathbf{R}_{\mathrm p}$.
\begin{equation}
T_\mathrm{pix}
\xrightarrow{\mathbf{R}_{\mathrm p}}
T(\theta_i, \phi_j, \boldsymbol{r}_{\mathrm p}).
\end{equation}
For a detailed description, see \appref{app:Coordinate transformation}.
The beam-convolved temperature $\tilde{T}(\boldsymbol{r}_{\mathrm p})$ is calculated by direct summation over the spherical beam grid.
\begin{equation}
\tilde{T}(\boldsymbol{r}_{\mathrm p})
=
\frac{
\displaystyle
\sum_{i,j}
B(\theta_i,\phi_j)
T(\theta_i,\phi_j;\boldsymbol{r}_{\mathrm p})
\sin\theta_i\,\Delta\theta\,\Delta\phi
}{
\displaystyle
\sum_{i,j}
B(\theta_i,\phi_j)
\sin\theta_i\,\Delta\theta\,\Delta\phi
}.
\end{equation}
Here, $\sin\theta_i\,\Delta\theta\,\Delta\phi$ is the solid angle represented by each sampling point.
The summation is performed over $0\leq\theta\leq\theta_{\mathrm{max}}$ and $0\leq\phi<2\pi$, with $\Delta\theta=0.1^\circ$ and $\Delta\phi=0.5^\circ$.
Since the beam is normalized as in \autoref{eq:beam_normalization}, the convolution at frequency $\nu$ reduces to
\begin{equation}
\tilde{T}_\nu(\boldsymbol{r}_{\mathrm p})
=
\frac{1}{4\pi}
\sum_{i,j}
B_\nu(\theta_i,\phi_j)
T_\nu(\theta_i,\phi_j;\boldsymbol{r}_{\mathrm p})
\sin\theta_i\,\Delta\theta\,\Delta\phi.
\label{eq:convolution}
\end{equation}
The choice of $\theta_{\mathrm{max}}$ has a negligible effect on the normalization because the total beam area is dominated by the main lobe.
It affects the chromaticity of the beam response, which we study later in \secref{subsec:sidelobes}.

After convolution, the resulting temperature $\tilde{T}(\boldsymbol{r}_{\mathrm p})$ is assigned to the WCS pixel containing $\boldsymbol{r}_{\mathrm p}$, so that the simulated maps have the same format as the observational data.
The simulated WCS maps are then interpolated onto the frequency grid of the real data to account for the difference in frequency resolution between the PO model ($1\,\mathrm{MHz}$) and the observations ($0.209\,\mathrm{MHz}$).
An example of the resulting simulated sky map is shown in \autoref{fig:map_sim}.

\section{Properties of the systematics}
\label{section:Properties of the systematics}
In the previous two sections, we described the observational data, simulation framework, and analysis methods adopted in this work.
We now turn to the observed L-band data and perform a series of diagnostic tests to investigate the origin and properties of the oscillatory systematics.
The insights gained from these tests motivate the design of our subsequent simulations and provide guidance for developing strategies to mitigate these systematics.

\begin{figure}
    \centering
    \includegraphics[width=1\linewidth]{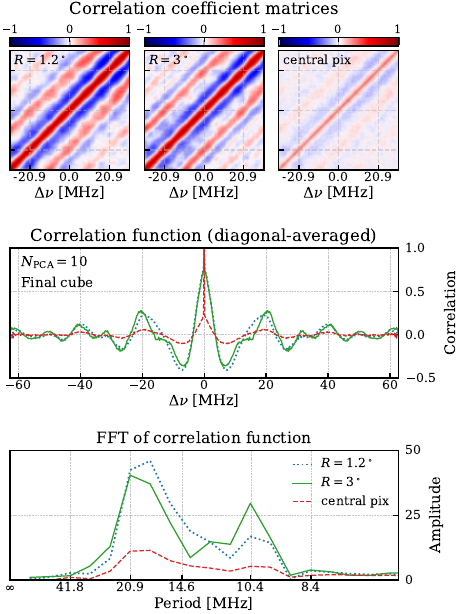}
    \caption{ Upper panel: correlation matrices of the cleaned final cube with $R_\mathrm{stack} = 0^\circ$ (left),  $R_\mathrm{stack} = 1.2 ^\circ$ (middle) and $R_\mathrm{stack} = 3.0^\circ$ (right).
    In all three cases $N_\mathrm{PCA} = 10$.
    Middle panel: the normalized two-point correlation functions obtained by averaging the matrices along their diagonals. 
    Lower panel: the Fourier transforms of the corresponding correlation functions, showing the characteristic period of the oscillating structure.
    The amplitude of the oscillating component depends on the stacking radius.}
    \label{fig:matrix_radius}
\end{figure}

\begin{figure}
    \centering
    \includegraphics[width=1\linewidth]{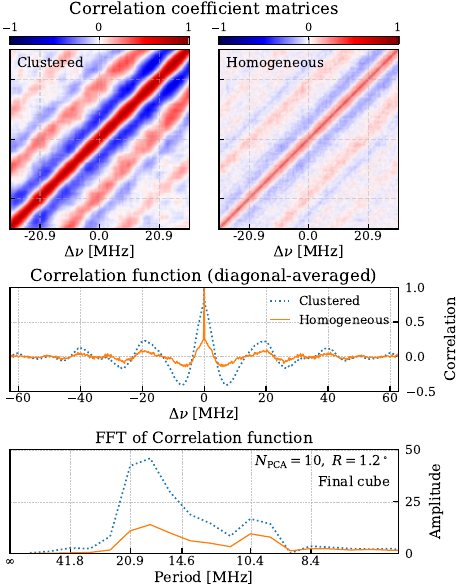}
    \caption{Upper panel: correlation matrices of the final cube with clustered mock galaxies (left) and homogeneous mock galaxies (right).
    In both cases $N_\mathrm{PCA} = 10$.
    Clustered galaxies enhance the ripple pattern.
    Middle panel: the normalized two-point correlation functions obtained by averaging the matrices along their diagonals. 
    Lower panel: the Fourier transforms of the corresponding correlation functions, showing the characteristic period of the oscillating structure.
    }
    \label{fig:matrix_no_clustering}
\end{figure}

\begin{figure*}
    \centering
    \includegraphics[width=1\linewidth]{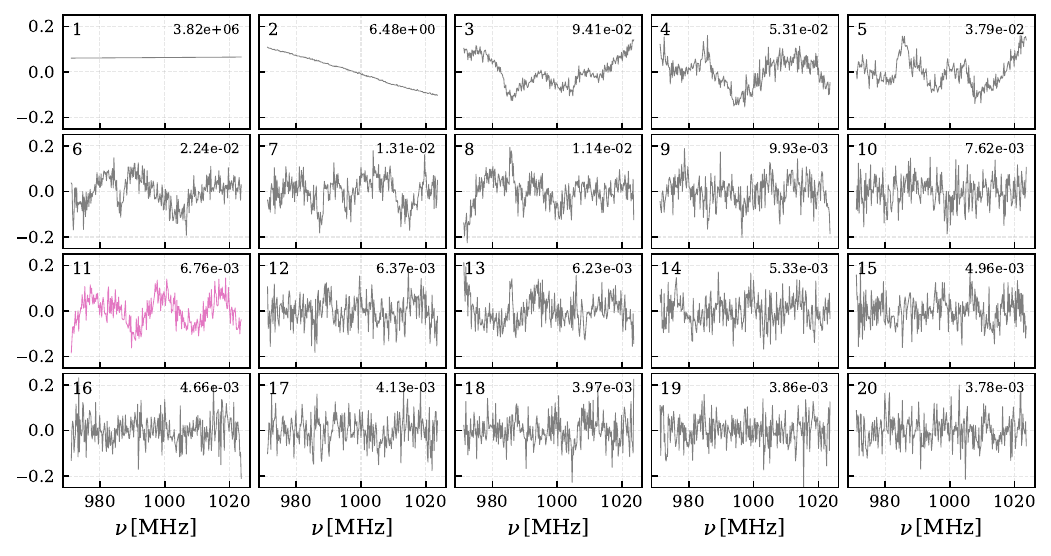}
    \caption{The first 20 PCA modes derived from the final MeerKLASS L-band deep-field cube. The eigenvalue associated with each mode is indicated in the upper right corner of each panel. The y-axis shows the dimensionless mode amplitude.}
    \label{fig:PCA_final_cube}
\end{figure*}
\subsection{Additive or multiplicative nature of systematics}
In PCA, the frequency spectrum of each pixel is regarded as a vector $\boldsymbol{v}$ which consists of three components, namely the foreground $\boldsymbol{v}_f$, the noise $\boldsymbol{v}_n$, and the HI signal $\boldsymbol{v}_{\mathrm{\hi}}$.
In our case, since the HI signal is much weaker than the other two components and we aim to investigate the systematics, we treat $\boldsymbol{v}_{\mathrm{\hi}}$ as part of $\boldsymbol{v}_{n}$.
\begin{equation}
    \boldsymbol{v} = \boldsymbol{v}_n+\boldsymbol{v}_f+\boldsymbol{v}_{\rm \hi} \approx  \boldsymbol{v}_n+\boldsymbol{v}_f\label{eq:vector combination}
\end{equation}
The dimension of $\boldsymbol{v}$ is the length of the frequency spectrum, which is 252 in this work.

PCA removes the foreground by finding a low-dimensional subspace $U$ and subtracting its projection on $U$ from $\boldsymbol{v}$ \citep{Spinelli2026}.
The orthonormal base vectors of $U$ are called PCA modes which are denoted by $\boldsymbol{u}_i$.
The dimension of $U$ can therefore be denoted as $N_{\mathrm{PCA}}$.
For the data analysis in \citet{Chen2025} and \citet{MeerKLASS2025}, $N_{\mathrm{PCA}}=10$ is used.
The projection of $\boldsymbol{v}$ is
\begin{equation}
\mathrm{P}_U(\boldsymbol{v}) = \sum_{i=1}^{N_\mathrm{PCA}}
(\boldsymbol{v} \cdot \boldsymbol{u}_i)\boldsymbol{u}_i.
\end{equation}
With \autoref{eq:vector combination}, we have
\begin{equation}
    \mathrm{P}_U(\boldsymbol{v}) = \mathrm{P}_U(\boldsymbol{v}_n)+\mathrm{P}_U(\boldsymbol{v}_f).
\end{equation}
The cleaned spectrum is 
\begin{equation}
    \boldsymbol{v}_c = \boldsymbol{v} - \mathrm{P}_U(\boldsymbol{v}) = [\mathbf{I}-\mathrm{P}_U](\boldsymbol{v}_n) + \left[ \boldsymbol{v}_f - \mathrm{P}_U(\boldsymbol{v}_f) \right],
\end{equation}
where $\mathbf{I}$ is the identity matrix.
The first term on the right-hand side represents the thermal noise plus \HI\ signal component, which is modulated by the projector $P_U$.
The second term is the foreground residual.
An ideal foreground removal process should preserve $\boldsymbol{v}_n$ with minimal distortion while removing $\boldsymbol{v}_f$ as much as possible.
Therefore, $[\mathbf{I}-P_U]$ leads to \emph{multiplicative systematics}, whereas $\left[ \boldsymbol{v}_f - \mathrm{P}_U(\boldsymbol{v}_f) \right]$ introduces \emph{additive systematics}.
 
To distinguish between these two systematics, we derive $P_U$ from the real final cube and apply it to a mock noise cube which is generated based on the real hit counts of the corresponding map pixels. The system temperature is assumed to be uniform at $T_\mathrm{sys}=17\,\mathrm{K}$ according to \citet{MeerKLASS2025}, due to the narrow frequency sub-band used for data analysis.
If the additive term is negligible, the correlation matrix derived from the noise cube should closely resemble that from the final cube since the noise covariance dominates in the residual after PCA cleaning.
However, they exhibit substantial differences, as shown in the upper panel of \autoref{fig:matrix_noise}.
The ripple pattern in the noise cube matrix is significantly weaker than that in the final cube matrix.
To facilitate the comparison, we further average each matrix along its diagonals to obtain a normalized two-point correlation function, hereafter referred to simply as the \emph{correlation function}, as shown in the middle panel of \autoref{fig:matrix_noise}.
Since the correlation function exhibits a clear oscillating structure, we compute its Fourier transform to identify the characteristic periods of the oscillation.
Hereafter, the Fourier transform of the correlation function is referred to as the \emph{Fourier spectrum}, as shown in the lower panel of \autoref{fig:matrix_noise}.
These comparisons validate that multiplicative effects induced by PCA are not sufficient to explain the systematics and the oscillating structure in the data is a kind of foreground residual.
Later in \secref{section:Simulation Results}, we demonstrate how the frequency-smooth foregrounds become oscillatory after convolution with the chromatic beam.

\subsection{Common features of systematics in adjacent pixels}
\label{section:Common feature of systematics in adjacent pixels}
The stacked/reference spectrum is constructed by summing spectra from multiple pixels.
A feature that remains after stacking must therefore be shared, at least to some extent, by the contributing pixels.
Such a common feature would normally be expected to be identified and removed by PCA.
However, the oscillatory structure persists after PCA cleaning and becomes prominent in the stacked/reference spectrum.
In this subsection, we propose a possible explanation and test it with the L-band final cube.

As discussed in the previous sections, the oscillatory structure comes from the convolution of the foreground with the telescope beam.
It is natural to expect this oscillatory structure to vary gradually with telescope pointing, in both amplitude and spectral shape.
PCA treats all spatial pixels equally, irrespective of their relative positions.
Across the full survey area, the oscillatory structure can vary substantially from pixel to pixel.
As a consequence, PCA finds no common feature among all of the pixels.
The stacked signals, on the other hand, integrate over pixels around the positions of the stacked galaxies.
Between neighbouring pixels, the oscillatory structure varies little. 
Summing these pixels accentuates the oscillation instead of cancelling them out.
In addition, the (mock) galaxies are clustered rather than uniformly distributed.
The oscillatory structures around galaxies which are closer tend to be more similar.
Together, these two effects cause stacking to preferentially combine a subset of pixels with relatively similar oscillatory features.
Instead of averaging out, these features can consequently become more prominent in the stacked/reference spectrum.

If this interpretation is correct, reducing either the contribution from neighbouring pixels or the spatial clustering of the galaxy sample should weaken the oscillatory structure.
We test these two effects separately by modifying the stacking procedure in two ways.
First, we remove the summation over neighbouring pixels and use only the central pixel at each galaxy position.
As shown in \autoref{fig:matrix_radius}, the ripple pattern in the correlation matrix is substantially reduced.
Increasing $R_\mathrm{stack}$, on the other hand, makes little difference.
Second, we remove the clustering of the mock galaxies but retain the neighbouring-pixel summation with $R_\mathrm{stack}=1.2^\circ$.
The ripple pattern is again substantially reduced, as shown in \autoref{fig:matrix_no_clustering}.
While not shown, we also validate that when both neighbouring-pixel summation and galaxy clustering are removed, the ripple pattern is further weakened.
These tests support the picture described above and the assumption that the oscillatory foreground residual varies with telescope pointing.
Similar phenomena are reproduced in our simulations (see \appref{app:Common feature in adjacent pixels}).

\subsection{Oscillation in PCA mode}
\label{section:Oscillation in PCA mode}
The choice of $N_\mathrm{PCA}$ is important for investigating residual systematics, as removing additional modes suppresses more correlated structures at the cost of increased \HI\ signal loss.
In \citet{MeerKLASS2025}, the first 10 PCA modes are removed ($N_\mathrm{PCA}=10$), as this provides a good balance between foreground removal and \HI\ signal loss.
\citet{Chen2025} adopted the same choice of $N_\mathrm{PCA}=10$, but did not perform the reconvolution step used in \citet{MeerKLASS2025}.
It is found that the reconvolution does not reduce the systematics in the stacked signal and instead makes the oscillations less localised in frequency. 
See the full discussion in \citet{Chen2025}.

\begin{figure}
    \centering
    \includegraphics[width=1\linewidth]{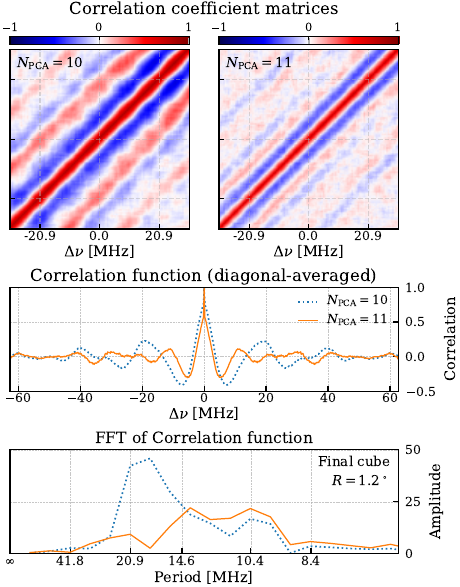}
    \caption{ Upper panel: correlation matrices of the final cube with $N_\mathrm{PCA} = 10$ (left) and $N_\mathrm{PCA} = 11$ (right).
    Middle panel: the normalized two-point correlation functions obtained by averaging the matrices along their diagonals. 
    Lower panel: the Fourier transforms of the corresponding correlation functions, showing the characteristic period of the oscillating structure.
    The oscillating component with a period of 20\,MHz is greatly suppressed after removing the 11th PCA mode.}
    \label{fig:matrix_final_cube_11}
\end{figure}

In this subsection, we focus on the properties of the residual systematics, while the associated \HI\ signal loss is discussed later.
In \autoref{fig:PCA_final_cube}, mode 11 exhibits a clear oscillating structure.
Therefore, we compare the correlation matrices for $N_\mathrm{PCA}=10$ and $11$.
\autoref{fig:matrix_final_cube_11} shows that removing mode 11 significantly weakens the ripple pattern in the correlation matrix, particularly the 20\,MHz component shown in the lower panel.
This confirms that the data cleaned with $N_\mathrm{PCA}=10$, as used in the previous power-spectrum and stacking analyses, contain at least one oscillating systematic component with a characteristic period of approximately $20\,\mathrm{MHz}$.
However, the ripple pattern does not disappear completely for $N_\mathrm{PCA}=11$, suggesting the presence of additional systematics that cannot be removed by simply subtracting one more PCA mode.
Later we shall see that removing 11 modes causes greater signal loss and does not improve the stacked spectrum.

\begin{figure*}
    \centering
    \includegraphics[width=\linewidth]{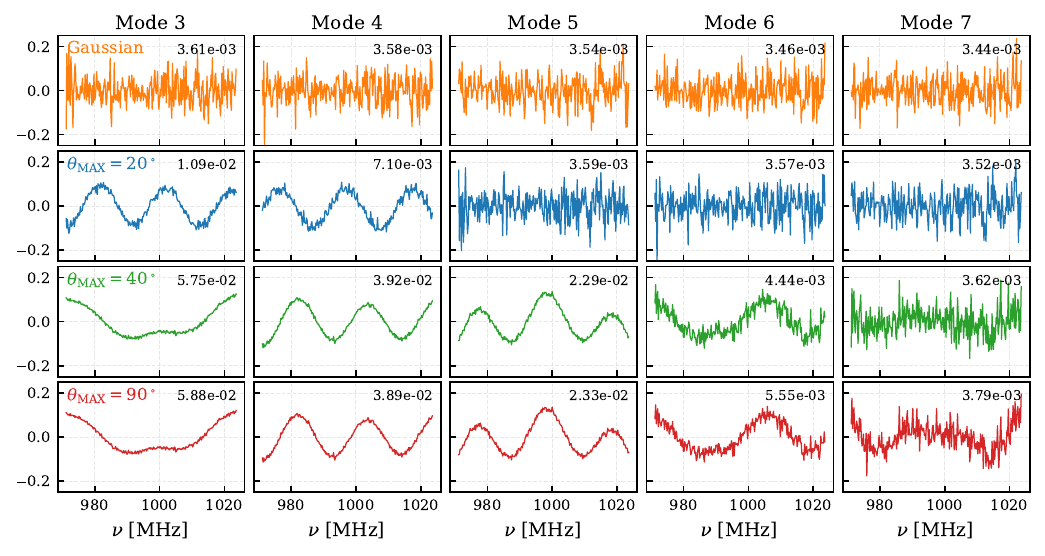}
    \caption{PCA modes derived from simulated data using different beam models. 
    The first row corresponds to a Gaussian beam with $\mathrm{FWHM}=1.02\lambda/D$, while the remaining rows show the results obtained with the PO beam truncated at different values of $\theta_{\mathrm{max}}$. 
    The eigenvalue associated with each mode is indicated in the upper right corner of each panel, and the y-axis shows the dimensionless mode amplitude. 
    The system temperature is set to $T_{\mathrm{sys}}=\sqrt{2}\times17\,\mathrm{K}$. 
    In each of the last three rows, a pair of nearly sinusoidal modes is identified. Together, they form a basis for representing a 20\,MHz sinusoidal oscillation with an arbitrary phase, indicating that they capture the oscillation introduced by the chromatic response of the main beam. As $\theta_\mathrm{max}$ increases, more structures emerge in the PCA modes.}
    \label{fig:PCA_sidelobe}
\end{figure*}

\begin{figure*}
    \centering
    \includegraphics[width=1\linewidth]{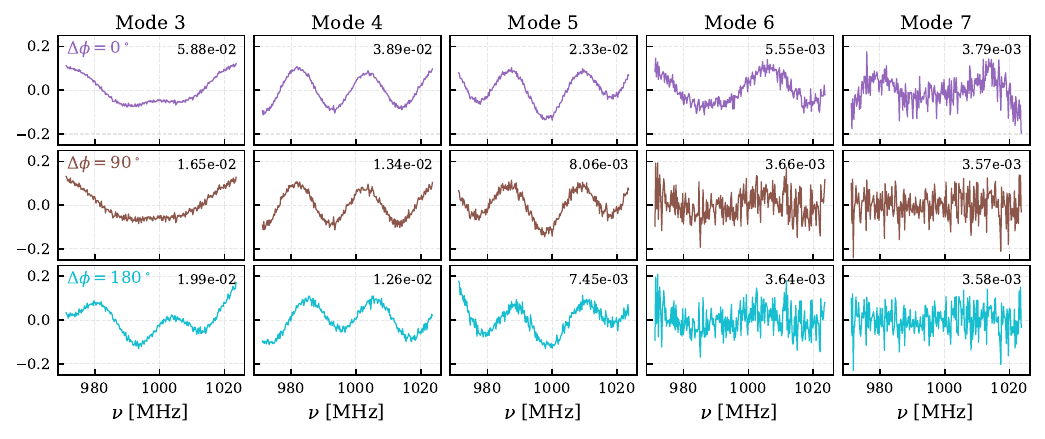}
    \caption{PCA modes derived from simulated data with different values of $\Delta\phi$. 
    The eigenvalue associated with each mode is indicated in the upper right corner of each panel, and the y-axis shows the dimensionless mode amplitude. 
    The system temperature is set to $T_{\mathrm{sys}}=\sqrt{2}\times17\,\mathrm{K}$ in the first row and $T_{\mathrm{sys}}=17\,\mathrm{K}$ in the last two rows, as the latter are averaged over two scans. 
    In the last two rows, the structures in modes~6 and 7 are not as clear as those in the first row.}
    \label{fig:PCA_rotation}
\end{figure*}

\begin{figure}
    \centering
    \includegraphics[width=1\linewidth]{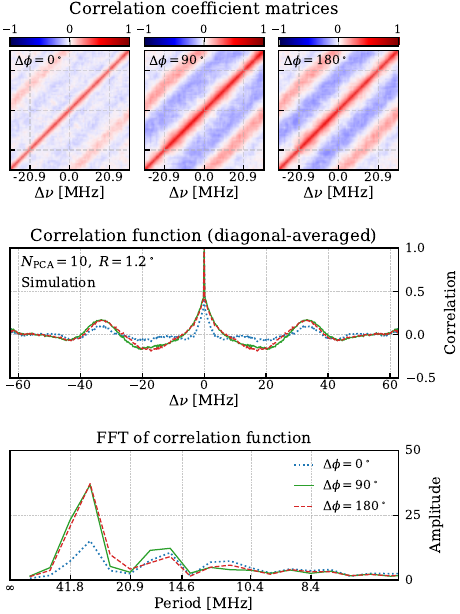}
    \caption{Upper panel: correlation matrices obtained from a single scan (left), the average of two scans with a beam rotation of $\Delta\phi=90^\circ$ (middle) and $180^\circ$ (right). 
    In all three cases, $N_{\mathrm{PCA}}=10$. 
    Middle panel: normalized two-point correlation functions obtained by averaging the corresponding correlation matrices along their diagonals. 
    Lower panel: Fourier transforms of the corresponding correlation functions, showing the characteristic period of the oscillating structure. 
    The oscillating structure is significantly enhanced after averaging the two scans, indicating that the increased spectral complexity degrades the performance of PCA cleaning.}
    \label{fig:matrix_rotation}
\end{figure}

\begin{figure}
    \centering
    \includegraphics[width=1\linewidth]{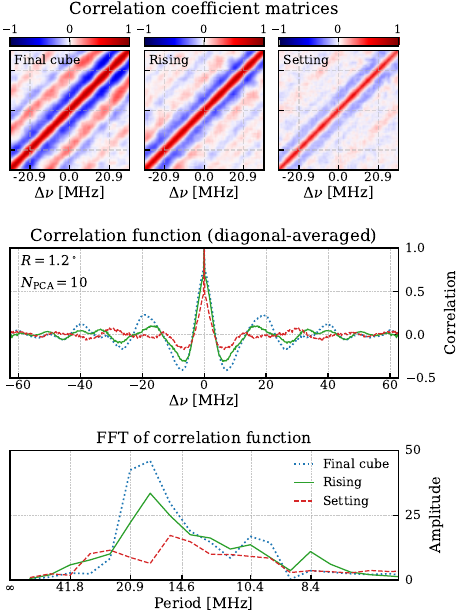}
    \caption{Upper panel: correlation matrices obtained from the rising cube (left), the setting cube (middle), and the final cube (right). 
    In all three cases, $N_{\mathrm{PCA}}=10$. 
    Middle panel: normalized two-point correlation functions obtained by averaging the corresponding correlation matrices along their diagonals. 
    Lower panel: Fourier transforms of the corresponding correlation functions, highlighting the characteristic period of the oscillating structure. 
    The oscillation is significantly stronger in the final cube than in either the rising or setting cubes, indicating that combining the two scanning strategies enhances the frequency-correlated systematics.}
    \label{fig:matrix_rising_setting}
\end{figure}

\section{Simulation results}
\label{section:Simulation Results}
In the previous section, we have established that the oscillating systematics are an additive residual component in the cleaned map. 
They are spatially clustered and not fully removed even with large numbers of PCA modes $N_{\rm PCA}>10$. 
The findings hint towards the connection between the systematics and the complicated chromatic structures of the MeerKAT beam, which we validate in this section using the full-sky PO beam model with the foreground simulation discussed in \secref{section:Simulation}.
We emphasise again that we only simulate the beam-convolved foregrounds and the thermal noise, and use the frequency correlation matrix as the indicator of systematics.
Since we have shown the systematics to be an additive component, the impact of \HI\ signal on the systematics is negligible.

\subsection{Far sidelobes}
\label{subsec:sidelobes}
In general, sidelobes are much weaker than the main lobe, and their effects are therefore often neglected, partly because of the lack of suitable tools to investigate them.
With our full-sky PO beam model, we first investigate how far from the beam centre the sidelobes can still have a non-negligible impact on the frequency spectrum.
We set an upper limit $\theta_{\mathrm{max}}$ on the angular extent of the beam model, such that sidelobes at $\theta > \theta_{\mathrm{max}}$ from the beam centre are excluded.
We consider $\theta_{\mathrm{max}} = 20^\circ$, $40^\circ$, and $90^\circ$.
\autoref{fig:PCA_sidelobe} shows the PCA modes for different cases, ordered by their corresponding eigenvalues from largest to smallest.
We start from mode~3 because the first mode is constant and the second mode is nearly linear; together, they represent the average frequency spectrum and are dominated by the smooth synchrotron structure.
From mode~8 onward, all modes appear to be dominated by noise.
For clarity, we do not include them in the comparison.
When $\theta_\mathrm{max} = 20^\circ$, only modes~3 and~4 show large-scale structures, and both look sinusoidal.
When $\theta_\mathrm{max} = 40^\circ$, two additional large-scale modes emerge.
The two sinusoidal modes are still present, but they shift to modes~4 and~5.
Mode~3 is replaced by a bowl-shaped curve, and mode~6 is no longer noise-like.
When $\theta_\mathrm{max} = 90^\circ$, even mode 7 shows a W-shaped structure.
For comparison, the first row of \autoref{fig:PCA_sidelobe} shows the corresponding PCA modes obtained with a Gaussian beam model of FWHM $1.02\lambda/D$.
None of these structures appears in the Gaussian beam case, suggesting that the Gaussian beam model is overly simplified.

These results provide two important insights.
First, the far sidelobes introduce additional spectral structures into the frequency spectra.
Second, what PCA captures is the foreground emission modulated by the chromatic beam response.
Together, these results help explain why the oscillatory structure in the frequency spectrum varies with telescope pointing as discussed in \secref{section:Common feature of systematics in adjacent pixels}. 
From \autoref{eq:convolution}, the measured frequency spectrum can be approximated as a weighted sum of the main lobe and all sidelobes,
\begin{equation}
\tilde{T}(\nu,\boldsymbol{r}_{\mathrm p})
\approx
\sum_{\boldsymbol{n}} B_{\boldsymbol{n}}(\nu)\,\mathcal{T}_{\boldsymbol{n}}(\boldsymbol{r}_{\mathrm p}),
\label{eq:linear_combination}
\end{equation}
where $\boldsymbol{n}$ denotes the direction relative to the beam centre, i.e. $(\theta,\phi)$.
The weight $\mathcal{T}_{\boldsymbol{n}}$ is proportional to the sky temperature seen in that direction and therefore varies with the telescope pointing $\boldsymbol{r}_{\mathrm p}$.
Since $B_{\boldsymbol{n}}(\nu)$ has a very different frequency dependence for different $\boldsymbol{n}$, changing these weights alters the relative contributions of the sidelobes \citep{Cerfonteyn2024}.
Consequently, the spectral shape of $\tilde{T}(\nu,\boldsymbol{r}_{\mathrm p})$ varies with the telescope pointing.

It is noteworthy that some of the modes show surprising similarities to those from the real data.
For example, in \autoref{fig:PCA_final_cube}, mode~3 has a bowl-shaped profile, and mode~6 resembles its counterpart in the simulation with $\theta_\mathrm{max}=90^\circ$, except for a sharp trough near 986\,MHz.
The sinusoidal mode~11 in the real data, which contributes strongly to the ripple pattern, resembles mode~5 for $\theta_\mathrm{max} \geq 40^\circ$ and mode~4 for $\theta_\mathrm{max} = 20^\circ$ in \autoref{fig:PCA_sidelobe}.

The simulations presented here represent an idealised description of the beam-induced effects.
In real observations, the effective beam response depends on the telescope pointing and scanning strategy and may be further affected by instrumental and environmental systematics.
In particular, for observations away from the zenith, some far sidelobes may extend below the local horizon and therefore couple predominantly to terrestrial emission rather than the sky considered in our simulations.
Ground-based RFI could consequently introduce additional spectral structures through these sidelobes.
Mutual coupling between nearby antennas may also modify the effective beam response and introduce additional frequency-dependent structures.
These effects may partly account for features present in the PCA modes of the real data but absent from the simulations.
Modelling such observational complexities is beyond the scope of this work and will be investigated in future studies.

\subsection{Beam rotation}
As mentioned in \secref{section:Stacking procedure with MeerKLASS L-band data}, the beam rotates around its centre between different scans.
If the beam orientations (i.e. the beam rotation angles) are uniformly or randomly distributed, the beam asymmetry will be averaged out, resulting in a simpler and more symmetric effective beam, as assumed by \citet{Matshawule2021}.
However, if the beam orientations remain close to one value during rising scans and another during setting scans, the asymmetry will not be averaged out.
Instead, the effective beam after combining these two scans will retain a more complex angular structure.

We model the second scenario in our simulations.
Specifically, we convolve the sky with the original beam and with a copy of the beam rotated by an angle $\Delta\phi$, producing two sky maps.
These two maps are then averaged to form the final simulated map, from which the correlation matrix is computed.
The PCA modes for $\Delta\phi = 0^\circ$, $90^\circ$, and $180^\circ$ are presented in \autoref{fig:PCA_rotation}.
The shape of mode~3 changes significantly with $\Delta\phi$, but its large-scale structure remains visible in all three cases.
However, the large-scale structures in modes~6 and~7 become less prominent.
One might expect that combining scans with different $\Delta\phi$ would produce a simpler effective beam, thereby mitigating the impact of beam chromaticity.
Instead, our results suggest the opposite: it makes the effective beam more complex and consequently makes PCA less effective.
This interpretation is inspired by the correlation matrices.
\autoref{fig:matrix_rotation} shows the correlation matrices with (1) no beam rotation, (2) beam rotation of $\Delta\phi = 90^\circ$ and (3) beam rotation of $180^\circ$, respectively.
The ripple patterns are substantially stronger when beam rotation is included than when no rotation is applied.
This suggests that more frequency-correlated systematics remain in the cleaned data when datasets observed with different beam orientations (i.e. original or rotated) are combined.
A possible explanation is that different beam orientations imprint different systematic structures on the data.
When each dataset is considered separately, these structures can appear as well-defined higher-order PCA modes, as shown in \autoref{fig:PCA_rotation}.
However, after datasets with different beam orientations are combined, the different systematic components are superimposed.
The variance associated with the systematics is therefore distributed among a broader set of directions in the eigenspace, rather than being concentrated in a small number of dominant modes.
Consequently, a fixed number of PCA modes may no longer capture these systematics as efficiently as when each dataset is cleaned separately.
The remaining frequency-correlated residuals then give rise to the stronger ripple pattern seen in the stacked spectrum and its estimated correlation matrix.
A simple ideal geometric interpretation of this effect is presented in \appref{app:Combination of two datasets}.
In the real observations, the target field is scanned by 64 dishes in 41 nights.
The impact of such a combination is more complicated, which may contribute to the fact that we see systematics persisting in higher-order eigenmodes in the data.

\begin{figure*}
    \centering
    \includegraphics[width=1\linewidth]{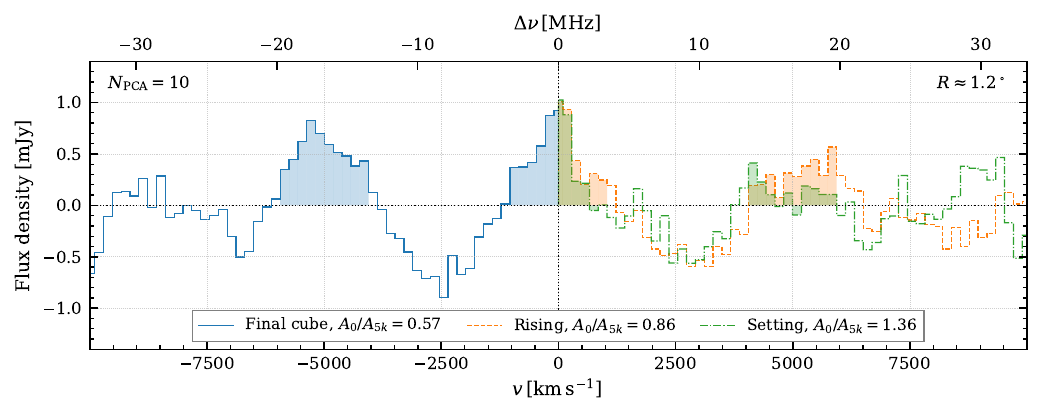}
    \caption{The symmetrised stacked spectra obtained using different data-splitting strategies.
    For clarity, only the negative-velocity half of the final-cube spectra is shown, while the positive-velocity half shows the results obtained by the rising cube only (orange line) and by the setting cube only (green line).
    $A_0$ and $A_{5k}$ are the areas of the shaded regions near $|v|=0$ km/s and $|v|= 5000$ km/s respectively.
    The central velocity channel is divided equally between the two halves of the figure.
    The area of the entire central peak is therefore $2A_0$.
    All spectra are integrated within an angular radius of $R_{\rm stack}\simeq1.2^\circ$.
    }
    \label{fig:stacked_spectrum_rising_setting}
\end{figure*}

\begin{figure*}
    \centering
    \includegraphics[width=1\linewidth]{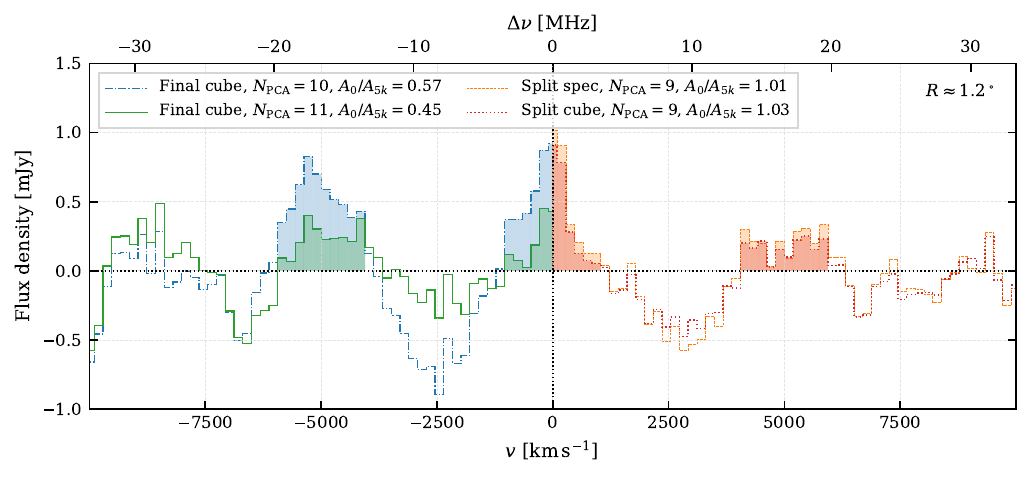}
    \caption{The symmetrised stacked spectra obtained using different data-splitting strategies.
    For clarity, only the negative-velocity half of the final-cube spectra is shown, while the positive-velocity half shows the results obtained by averaging split stacked spectra (orange solid line) and by averaging split cubes (dotted red line).
    $A_0$ and $A_{5k}$ are the areas near $|v|=0$ km/s and $|v|= 5000$ km/s respectively.
    The central velocity channel is divided equally between the two halves of the figure.
    All spectra are integrated within an angular radius of $R_{\rm stack}\simeq1.2^\circ$.
    $A_0/A_{5k}$ is about 1.6 times higher after splitting.
    }
    \label{fig:stacked_spectrum_improvement}
\end{figure*}

\section{Improved stacked spectrum with L-band data}
\label{section:Improved stacked spectrum with L-band data}
The simulations have shown that combining scans with two different beam orientations hinders PCA from capturing the systematics and leads to stronger frequency correlations after cleaning.
Therefore, we use the insights from the simulations to optimise the data analysis for the stacking measurement.

In \citet{Chen2025}, data from all scans were combined to construct the three-dimensional intensity map, i.e. the final cube $T(\mathrm{R.A.}, \mathrm{Dec.}, \nu)$.
10 PCA modes were removed from the final cube before calculating the reference spectrum, correlation matrix, and stacked spectrum.
Here, we adopt a different approach.
We construct the rising and setting cubes by separately combining the data from all rising and setting scans, respectively.
PCA foreground cleaning is then applied independently to the two cubes.
In \autoref{fig:matrix_rising_setting} and \autoref{fig:stacked_spectrum_rising_setting}, we remove 10 PCA modes from each cube to allow a direct comparison with the results obtained from the final cube.
In \autoref{fig:stacked_spectrum_improvement}, we instead remove 9 PCA modes from the rising and setting cubes, as this is the optimised case in which the ratio of the \HI\ signal to the foreground residual reaches its maximum.

\autoref{fig:matrix_rising_setting} shows that the ripple pattern is substantially weaker in the split data than in the final cube, with the greatest reduction occurring in the setting scans.
The correlation matrices mainly reflect the foreground residuals.
We therefore also examine the stacked spectrum.
In the right half of \autoref{fig:stacked_spectrum_rising_setting}, the stacked spectra of the rising and setting cubes are presented.
In the left half, we present the original result from \citet{Chen2025} which is computed from the final cube.
All stacked spectra have been symmetrised \citep{Sinigaglia2022,Chen2025} so there is no loss of information in only showing half of them.
The central peak is expected to contain the \HI\ emission, while the other structures in the stacked spectrum are mainly associated with the systematics.
Near $|v|=5000\,\mathrm{km\,s^{-1}}$, the oscillation on the right is significantly weaker than that on the left.

To quantify this difference, we compare the areas of the central peak and the side peaks near $|v|=5000\,\mathrm{km\,s^{-1}}$.
The trough near $|v|=2500\,\mathrm{km\,s^{-1}}$ is partly caused by the mean subtraction in the PCA cleaning, which forces the cleaned spectrum to have approximately zero mean.
Therefore, the positive excess around $v=0\,\mathrm{km\,s^{-1}}$ must be accompanied by a negative excess elsewhere, making the trough unsuitable for independently measuring the foreground residuals.
We therefore use the side peaks near $|v|=5000\,\mathrm{km\,s^{-1}}$.
Specifically, we select half of the central bin and the 5 adjacent bins ($\Delta v \approx 1036\,\mathrm{km\,s^{-1}}$), corresponding to approximately half of the full width of the central peak, and denote the integrated area over this range as $A_0$.
The area of the entire central peak should be $2A_0$.
For the side peaks, we select 10 bins $\Delta v\approx1884\,\mathrm{km\,s^{-1}}$ around $|v|=5000\,\mathrm{km\,s^{-1}}$ and denote the area by $A_{5k}$.
We use $A_0$ and $A_{5k}$ to characterise the amplitudes of signal and foreground residuals, respectively.
These regions are indicated by the shaded areas in \autoref{fig:stacked_spectrum_rising_setting}.
In the original result in \citet{Chen2025}, $2A_0/A_{5k} \approx1.14$, meaning that the oscillating systematics are almost as strong as the \HI\ emission.
For the rising and the setting cube, this ratio increases to $2A_0/A_{5k} \approx 1.72$ and $2.72$, respectively.
These results indicate that combining data from the rising and setting scans significantly enhances the foreground residual.
Additionally, the foreground residual in the setting data is much weaker than that in the rising data.

Although using only one type of scan can effectively mitigate the oscillating foreground residuals, only half of the observation time is used.
To make full use of the observations, we still need to combine the rising and setting scans, but after PCA cleaning.
We test two approaches, which give very similar results.
In the first approach, we directly average the two stacked spectra obtained from the rising and setting cubes.
This result is denoted as ``split spec'' in \autoref{fig:stacked_spectrum_improvement}.
In the second approach, we first apply PCA cleaning to the rising and setting cubes separately.
The two cleaned cubes are then combined using a weighted average.
For each voxel, we use the hit counts as the weight.
The stacked spectrum obtained from the combined cube is denoted as ``split cube'' in \autoref{fig:stacked_spectrum_improvement}.
As discussed in \secref{section:Oscillation in PCA mode}, increasing $N_{\mathrm{PCA}}$ reduces the foreground residuals but also leads to greater signal loss.
Therefore, it is generally desirable to use as few PCA modes as possible to capture the foreground.
By combining the observations after PCA cleaning, the oscillating foreground residuals are substantially reduced while only 9 PCA modes are removed, as shown in the right half of \autoref{fig:stacked_spectrum_improvement}.
Compared with the result of \citet{Chen2025}, shown by the blue dash-dot line in the left half of \autoref{fig:stacked_spectrum_improvement}, the central bin on the right is comparable in amplitude or even higher, depending on which approach is used.
The systematic component is reduced so that the ratio increases to $2A_0/A_{5k} \approx 2.0$ in both cases,
an increase of $81\%$ compared with the original result.
This demonstrates that our strategy suppresses the systematic effects while preserving the \HI\ signal, achieving both goals simultaneously.
In this sense, we alleviate the traditional trade-off in PCA foreground cleaning: removing too few modes leaves significant foreground residuals, while removing too many modes leads to substantial signal loss.

This can be further illustrated by the stacked spectrum obtained after removing 11 PCA modes from the final cube, as shown by the green solid line in the left half of \autoref{fig:stacked_spectrum_improvement}.
As discussed in \secref{section:Oscillation in PCA mode}, while removing mode 11 reduces the oscillations in the estimated correlation matrix,
\autoref{fig:stacked_spectrum_improvement} shows that this also reduces the amplitude of the \HI\ signal.
In this case, $2A_0/A_{5k}$ is only $0.9$, which is lower than the $N_\mathrm{PCA}=10$ case.

\section{Conclusion and discussion}
\label{section:Conclusion}
In this work, we illustrate the origin of the oscillating frequency-correlated systematics in the MeerKLASS L-band Deep-field data and develop a simple strategy to mitigate them.
We show that the oscillation originates from frequency-smooth foreground emission modulated by the chromatic MeerKAT beam.
After PCA cleaning, part of this beam-modulated foreground remains in the data and appears as a frequency-correlated residual.

This oscillation can be seen in the correlation matrix even after removing 10 PCA modes in the analysis in \citet{Chen2025}, and in this work we find a similar structure persists in mode 11. 
However, removing mode 11 causes greater \HI\ signal loss and does not improve the stacked spectrum, suggesting that the systematics are more broadly spread among the eigenmodes.

We also find that the oscillating residuals are spatially correlated.
The stacking procedure of \citet{Chen2025}, which uses clustered galaxy positions and sums the signal within $1.2^\circ$ around each galaxy, therefore enhances the visibility of these spatially correlated structures.
This phenomenon is also seen in the simulations, indicating that it arises naturally from the coupling between the angular structure of the foreground sky and the telescope beam.
Since standard PCA treats different lines of sight as independent samples, this spatial information is not explicitly exploited.
Including such information may provide an additional route for improving foreground cleaning, such as multi-scale PCA \citep{Carucci2025} or GMCA \citep{Carucci2020}.

To investigate the role of the beam, we simulate the observations using a full-sky electromagnetic model of the MeerKAT beam.
Unlike a simple Gaussian approximation, this model includes the complex frequency dependence and far sidelobes of the beam.
We find that a wide range of sidelobes at $\theta \lesssim 20^\circ,\,40^\circ,\,90^\circ$ can introduce additional oscillatory spectral structures into the foreground emission, even those at large angular distances of $40^\circ < \theta < 90^\circ$ from the beam centre.
These beam-induced structures can dominate some of the leading PCA modes and therefore affect foreground cleaning.

An additional complication arises from the rotation of the beam with respect to the sky.
The rising and setting scans observe the same field with different beam orientations.
Our simulations show that combining these orientations before PCA cleaning produces a more complex effective beam and strengthens the residual frequency correlations.
This explains why the systematic effect can become more difficult to remove when all observations are combined before foreground cleaning.
Future simulations and forecasts for \HI\ intensity-mapping surveys should therefore account not only for beam chromaticity and sidelobes, but also for changes in beam orientation during the observations.

Motivated by the simulations, we process the rising and setting observations separately.
In the real data, the frequency correlations in both subsets are substantially weaker than those in the final cube, confirming the behaviour predicted by the simulations.
We also find that the oscillation is considerably weaker in the setting data than in the rising data.
This difference may indicate contributions from additional observational systematics such as RFI, whose environment can vary with observing time.
A detailed investigation is beyond the scope of this work.
Since MeerKLASS observations involve multiple dishes and observing sessions, similar splits could be used in future work to identify which dishes, nights, or observing conditions show the strongest systematic effects and thereby help constrain their origin.

We apply PCA cleaning to the rising and setting cubes separately and combine them only after foreground subtraction.
With only 9 PCA modes removed instead of 10 in the previous analysis, this approach substantially suppresses the oscillating foreground residuals while preserving the stacked \HI\ signal.
The ratio of the \HI\ signal amplitude to the systematic amplitude increases by approximately $81\%$ compared with the original analysis.
Removing an additional PCA mode from the final cube can also suppress the oscillation, but at the cost of greater \HI\ signal loss.
Our strategy therefore alleviates the traditional trade-off in PCA foreground cleaning between foreground removal and signal preservation.

Finally, we comment on the implications of our findings for the power spectrum analysis using MeerKLASS data.
It is established in this work that the systematics are an additive foreground residual component of the data vector, which does not correlate with optical galaxy samples. 
Therefore, the cross-correlation with optical galaxies will remove the systematic contribution from the power spectrum measurements, while still having an impact on the covariance.
Our findings on the split data provide additional support for the recent auto-power spectrum measurements presented in \cite{MeerKLASS2026}.
By splitting the datasets to form the ``super-cross'' (see Fig.~2 of \citealt{MeerKLASS2026}) and cleaning the subsets separately,
the measurements are free of significant systematics as shown in the comprehensive validation tests therein.

More generally, our results demonstrate that accounting for the observing strategy and beam orientation can improve foreground cleaning without requiring more aggressive PCA subtraction, which sheds light on future \HI\ intensity-mapping analyses with MeerKAT and SKA-Mid.

\section*{Acknowledgements}
This work is supported by the National SKA Program of China (Nos. 2022SKA0110100, 2022SKA0110101, 2022SKA0110200, 2022SKA0110203), the NSFC International (Regional) Cooperation and Exchange Project (No.12361141814), the NSFC Innovation Group Project (No.12421003), the Specialized Research Fund for State Key Laboratory of Radio Astronomy and Technology, and the National Astronomical Observatories, Chinese Academy of Sciences (No. E5ZB0901).
B.Z. acknowledges the financial support from the China Scholarship Council (CSC; No. 202404910399).
ZC is supported by the Poincar\'e fellowship, and was funded by a UKRI Future Leaders Fellowship [grant MR/X005399/1; PI: Alkistis Pourtsidou].
DILdV was supported by the South African Radio Astronomy Observatory, which is a facility of the National Research Foundation, an agency of the Department of Science and Technology (Grant Number: 75322).
JF acknowledges support of Funda\c{c}\~{a}o para a Ci\^{e}ncia e a Tecnologia through the Scientific Employment Incentive program reference No.\ 2020.02633.CEECIND/CP1631/CT0002, the FCT exploratory project 10.54499/2023.15069.PEX, and the research grant UID/04434/2025.
Some of the results in this paper have been derived using the healpy and HEALPix packages.

\section*{Author contributions}
BZ is the lead author. 
The authors listed alphabetically from EdLA to DILdV make direct contributions to the results of this work. 
The authors listed alphabetically from MBS to LW are data builders for the MeerKLASS L-band intensity maps used in this work, or provided valuable insights that have impacted this work. 
EdLA and MGS conceived the project. 
BZ conducted the research work and led the paper write-up. 
EdLA, VM, and JC provided weekly scientific supervision throughout the project. 
ZC and MS provided regular scientific guidance and played a key role in the interpretation and analysis of the MeerKLASS data. 
DILdV provided the beam model for the MeerKAT dish and the idea for the physical mechanism of the pointing dependent beam chromaticity variations. 
JW provided the calibrated data. 
MBS provided the split data. 
FD resolved several programming issues. 
WH, XC and YW contributed valuable insights through scientific discussions. 
PB, SCa, IPC, SCu, BNE, JF, KG, MOI, and YL contributed to the building of the MeerKLASS L-band data.

\section*{Data Availability}

Placeholder for the data availability statement.



\bibliographystyle{mnras}
\bibliography{reference} 




\appendix


\section{Coordinate transformation}
\label{app:Coordinate transformation}
We use two orthonormal bases, corresponding to two right-handed coordinate systems, to describe directions on the sky.
The fixed sky basis is denoted by
\begin{equation}
\left\{
\boldsymbol{e}_{x'},
\boldsymbol{e}_{y'},
\boldsymbol{e}_{z'}
\right\},
\end{equation}
where $\boldsymbol{e}_{z'}$ points towards the North Celestial Pole.
The corresponding spherical coordinates are $(\theta',\phi')$, with $\theta'=90^\circ-\mathrm{Dec.}$ and $\phi'=\mathrm{R.A.}$.
For each beam pointing, we introduce a beam basis
\begin{equation}
\left\{
\boldsymbol{e}_{x},
\boldsymbol{e}_{y},
\boldsymbol{e}_{z}
\right\},
\end{equation}
where $\boldsymbol{e}_{z}$ is the pointing direction of the beam, also denoted as $\boldsymbol{r}_{\mathrm p}$.
The vector $\boldsymbol{e}_{x}$ lies in the plane spanned by $\boldsymbol{r}_{\mathrm p}$ and the North Celestial Pole and points towards the latter, while $\boldsymbol{e}_{y}$ completes the right-handed orthonormal basis.

Let $(\theta'_{\mathrm p},\phi'_{\mathrm p})$ be the spherical coordinates of the pointing direction in the sky coordinate system.
The pointing vector can then be expanded in the sky basis as
\begin{equation}
\boldsymbol{r}_{\mathrm p} = \boldsymbol{e}_{z}
=
\sin\theta'_{\mathrm p}\cos\phi'_{\mathrm p}\,\boldsymbol{e}_{x'}
+
\sin\theta'_{\mathrm p}\sin\phi'_{\mathrm p}\,\boldsymbol{e}_{y'}
+
\cos\theta'_{\mathrm p}\,\boldsymbol{e}_{z'}.
\label{eq:e_z}
\end{equation}
The beam $x$-basis vector is obtained by projecting the North Celestial Pole direction $\boldsymbol{e}_{z'}$ onto the plane perpendicular to $\boldsymbol{e}_{z}$.
After normalization, this gives
\begin{equation}
\boldsymbol{e}_{x}
=
-\cos\theta'_{\mathrm p}\cos\phi'_{\mathrm p}\,\boldsymbol{e}_{x'}
-\cos\theta'_{\mathrm p}\sin\phi'_{\mathrm p}\,\boldsymbol{e}_{y'}
+\sin\theta'_{\mathrm p}\,\boldsymbol{e}_{z'}.
\label{eq:e_x}
\end{equation}
Given that $\left\{\boldsymbol{e}_{x},\boldsymbol{e}_{y},\boldsymbol{e}_{z}\right\}$ is right-handed, we define
\begin{equation}
\boldsymbol{e}_{y}
\equiv
\boldsymbol{e}_{z}\times\boldsymbol{e}_{x}
=
\sin\phi'_{\mathrm p}\,\boldsymbol{e}_{x'}
-\cos\phi'_{\mathrm p}\,\boldsymbol{e}_{y'}.
\label{eq:e_y}
\end{equation}
Thus, all three beam basis vectors are expressed as linear combinations of the fixed sky basis vectors, with coefficients determined by the pointing direction.

Now consider an arbitrary direction vector $\boldsymbol{r}$ with spherical coordinates $(\theta,\phi)$ in the beam coordinate system.
In the beam basis, it can be written as
\begin{equation}
\boldsymbol{r}
=
\sin\theta\cos\phi\,\boldsymbol{e}_{x}
+
\sin\theta\sin\phi\,\boldsymbol{e}_{y}
+
\cos\theta\,\boldsymbol{e}_{z}.
\label{eq:r_in_beam_coordinate_system}
\end{equation}
It can also be expanded in the sky basis as
\begin{equation}
\boldsymbol{r}
=
x'\,\boldsymbol{e}_{x'}
+
y'\,\boldsymbol{e}_{y'}
+
z'\,\boldsymbol{e}_{z'}.
\label{eq:r_in_sky_coordinate_system}
\end{equation}
Substituting \autoref{eq:e_z}--\autoref{eq:e_y} into \autoref{eq:r_in_beam_coordinate_system} and comparing with \autoref{eq:r_in_sky_coordinate_system}, we get
\begin{equation}
\begin{bmatrix}
x'\\
y'\\
z'
\end{bmatrix}
=
\mathbf{R}_{\mathrm p}
\begin{bmatrix}
\sin\theta\cos\phi\\
\sin\theta\sin\phi\\
\cos\theta
\end{bmatrix},
\end{equation}
where
\begin{equation}
\mathbf{R}_{\mathrm p}
=
\begin{bmatrix}
-\cos\theta'_{\mathrm p}\cos\phi'_{\mathrm p}
&
\sin\phi'_{\mathrm p}
&
\sin\theta'_{\mathrm p}\cos\phi'_{\mathrm p}
\\
-\cos\theta'_{\mathrm p}\sin\phi'_{\mathrm p}
&
-\cos\phi'_{\mathrm p}
&
\sin\theta'_{\mathrm p}\sin\phi'_{\mathrm p}
\\
\sin\theta'_{\mathrm p}
&
0
&
\cos\theta'_{\mathrm p}
\end{bmatrix}.
\end{equation}
The three columns of $\mathbf{R}_{\mathrm p}$ are the coordinate components of $\boldsymbol{e}_{x}$, $\boldsymbol{e}_{y}$, and $\boldsymbol{e}_{z}$ in the sky basis, respectively.

The matrix $\mathbf{R}_{\mathrm p}$ therefore establishes a one-to-one mapping from the beam-grid points $(\theta,\phi)$ to the corresponding sky directions $(x',y', z')$.
The sky temperature at this direction is then interpolated from the HEALPix map and assigned to the beam-grid point, giving $T(\theta,\phi,\boldsymbol{r}_{\mathrm p})$.

When investigating the effect of beam rotation, we keep the definition of the beam basis unchanged.
Instead, a rotated beam is constructed by shifting the beam response in the $\phi$ direction.
The rotated beam $B_{\Delta\phi}$ is defined by
\begin{equation}
B_{\Delta\phi}(\theta,\phi)
=
B(\theta,\phi+\Delta\phi),
\end{equation}
where $\Delta\phi$ is the rotation angle.
The same coordinate transformation described above is then applied to the rotated beam.



\section{Toy models}
\subsection{Correlation introduced by PCA}
In this section, we build a toy model to illustrate that removing certain modes could leads to correlation in uncorrelated data.
Suppose we have some Gaussian noise spectra which can be seen as $N$-dimensional (column) vectors.
$\boldsymbol{n}$ represents one of them.
It's covariance matrix is 
\begin{equation}
    \left<\boldsymbol{n}\boldsymbol{n}^T \right>= \sigma^2I,
\end{equation}
where $\boldsymbol{n}^T$ is the transpose of $\boldsymbol{n}.$
Suppose we have two PCA modes which can be treated as two unit vectors.
One is called $\boldsymbol{c}$ and the other is called $\boldsymbol{s}$.
The $i$th components of $\boldsymbol{c}$ and $\boldsymbol{s}$, are defined as
\begin{equation}
    c_i = A \cos{k\nu_i},\qquad s_i =B \sin{k\nu_i}.
\end{equation}
respectively, where $\nu$ is frequency, $A,\ B$ and $k$ are arbitrary constants.
As PCA modes, $\boldsymbol{c}$ and $\boldsymbol{s}$ must be orthonormal.
\begin{equation}
    \boldsymbol{c}^{T}\boldsymbol{c}=1,\qquad
    \boldsymbol{s}^{T}\boldsymbol{s}=1,\qquad
    \boldsymbol{c}^{T}\boldsymbol{s}=0.
\end{equation}
This means
\begin{equation}
    A^2 = \left[ \sum_{i=1}^{N} \cos^2{k\nu_i}\right]^{-1} \approx \frac{2}{N},\qquad B^2 = \left[\sum_{i=1}^{N} \sin^2{k\nu_i}\right]^{-1}\approx \frac{2}{N}.
\end{equation}
If there are enough periods in the frequency range we cut, A and B be should be approximately the same.
Now let's do PCA.
First, we project $\boldsymbol{n}$ onto $\boldsymbol{c}$ and subtract the projected vector $\boldsymbol{c}\boldsymbol{c}^T\boldsymbol{n}$ from $\boldsymbol{n}$.
Then we repeat this process with $\boldsymbol{s}$. 
After this, the noise vector $\boldsymbol{n}$ becomes a new vector
\begin{equation}
    \tilde{\boldsymbol{n}} = (I -\boldsymbol{c}\boldsymbol{c}^T-\boldsymbol{s}\boldsymbol{s}^T)\boldsymbol{n}.
\end{equation}
For clarity, we define $P \equiv (I -\boldsymbol{c}\boldsymbol{c}^T-\boldsymbol{s}\boldsymbol{s}^T)$.
Obviously, $P = P^T$.
The covariance matrix of $\tilde{\boldsymbol{n}}$ is 
\begin{equation}
    \left<\tilde{\boldsymbol{n}}\tilde{\boldsymbol{n}}^T\right> = \left<P\boldsymbol{n} (P\boldsymbol{n})^T\right> = \left<P\boldsymbol{n}\boldsymbol{n}^TP^T \right>= \sigma^2PP^T = \sigma^2P^2.
\end{equation}
Since $P$ is an orthogonal projection matrix, 
\begin{equation}
    P^{2} = P
\end{equation}
which can also be proved by expanding
$(I-\boldsymbol{c}\boldsymbol{c}^{T}-\boldsymbol{s}\boldsymbol{s}^{T})
(I-\boldsymbol{c}\boldsymbol{c}^{T}-\boldsymbol{s}\boldsymbol{s}^{T})$, 
and using the orthonormality conditions.
The covariance matrix after PCA cleaning is
\begin{equation}
\left\langle
\tilde{\boldsymbol{n}}\tilde{\boldsymbol{n}}^{T}
\right\rangle
=
\sigma^2\mathbf{P}
=
\sigma^2
\left(
\mathbf{I}
-
\boldsymbol{c}\boldsymbol{c}^{T}
-
\boldsymbol{s}\boldsymbol{s}^{T}
\right).
\end{equation}
Its $(i,j)$ component is therefore
\begin{align}
\left\langle
\tilde{\boldsymbol{n}}\tilde{\boldsymbol{n}}^{T}
\right\rangle_{ij}
&=
\sigma^{2}\left[
\delta_{ij}
-
A^{2}\cos(k\nu_i)\cos(k\nu_j)
-
B^{2}\sin(k\nu_i)\sin(k\nu_j)
\right]
\nonumber\\
&=
\sigma^{2}\left[
\delta_{ij}
-
\frac{2}{N}\cos\bigl(k\nu_i-k\nu_j\bigr)
\right].
\end{align}
We have therefore shown that Gaussian noise can become correlated across frequency after sinusoidal PCA modes are removed.
Nevertheless, the induced correlation is weak, since the factor $2/N$ is of order $10^{-2}$ for spectra containing hundreds of frequency channels.

\subsection{Combination of two datasets}
\label{app:Combination of two datasets}
Let's consider two datasets in a two-dimensional vector space.
Each data point is represented by a vector $\mathbf{v}$.
We are interested in the case where the data within each dataset are highly correlated.
Therefore, within each dataset, we assume that the two components of the vectors follow independent Gaussian distributions.
Their joint distribution forms an ellipse in the two-dimensional vector space.

Let $\mathbf{u}$ be a unit vector.
The projection of a data vector $\mathbf{v}$ along $\mathbf{u}$ is $\mathbf{v}\cdot\mathbf{u}$.
For a set of data vectors, the projected values generally vary from one sample to another.
The variance of these projected values quantifies the magnitude of this variation along $\mathbf{u}$ and is defined as
\begin{equation}
\sigma_{\mathbf{u}}^2
\equiv
\left\langle
\left[
(\mathbf{v}-\langle\mathbf{v}\rangle)\cdot\mathbf{u}
\right]^2
\right\rangle,
\end{equation}
where $\langle\cdot\rangle$ denotes the average over all data vectors.
More generally, we may consider two different directions, $\mathbf{u}$ and $\mathbf{w}$, and ask how the corresponding projected fluctuations are correlated.
This leads to the covariance
\begin{equation}
C_{\mathbf{u},\mathbf{w}}
\equiv
\left\langle
\left[
(\mathbf{v}-\langle\mathbf{v}\rangle)\cdot\mathbf{u}
\right]
\left[
(\mathbf{v}-\langle\mathbf{v}\rangle)\cdot\mathbf{w}
\right]
\right\rangle.
\end{equation}
The covariance $C$ is a map that takes two vectors and returns a scalar, i.e. a symmetric $(0,2)$ tensor \citep{Wald1984}.

We choose the orthonormal basis $\{\mathbf{e}_x,\mathbf{e}_y\}$, where $\mathbf{e}_x$ and $\mathbf{e}_y$ point along the major and minor axes of the ellipse formed by dataset 1, respectively.
Once the basis is chosen, the components of $C$ form a matrix known as correlation matrix.
We denote the covariance matrix of dataset 1 by $\mathbf{C}_1$.
Since the components of the vectors along $\mathbf{e}_x$ and $\mathbf{e}_y$ follow independent Gaussian distributions $\mathbf{C}_1$ is diagonal,
\begin{equation}
\mathbf{C}_1
=
\begin{bmatrix}
\lambda & 0\\
0 & \epsilon
\end{bmatrix},
\qquad
\lambda\gg\epsilon.
\end{equation}
In this case, the variance along the major axis is given by the larger eigenvalue, $\lambda$, while the variance along the minor axis is given by the smaller eigenvalue, $\epsilon$.
The axis ratio scales as
\begin{equation}
\frac{L_x}{L_y}\propto\sqrt{\frac{\lambda}{\epsilon}}.
\end{equation}
We now consider dataset 2.
We assume that dataset 2 has the same elliptical shape as dataset 1, but is rotated by an angle $\alpha$.
The rotation is represented by
\begin{equation}
\mathbf{R}_{\alpha}
=
\begin{bmatrix}
\cos\alpha & -\sin\alpha\\
\sin\alpha & \cos\alpha
\end{bmatrix}.
\end{equation}
The covariance matrix of dataset 2 is therefore
\begin{equation}
\mathbf{C}_2
=
\mathbf{R}_{\alpha}
\mathbf{C}_1
\mathbf{R}_{\alpha}^{-1}.
\end{equation}

Thus, the two datasets have the same major and minor axis lengths, but their axes point in different directions.
We now combine the two datasets by averaging their corresponding vectors,
\begin{equation}
\mathbf{v}_{\mathrm c}
=
\frac{\mathbf{v}_1+\mathbf{v}_2}{2}.
\end{equation}
Apart from an overall factor that changes only the size of the ellipse, the covariance operator of the combined distribution is
\begin{equation}
\mathbf{C}
\propto
\mathbf{C}_1+\mathbf{C}_2.
\end{equation}
The two eigenvalues of $\mathbf{C}$ are
\begin{equation}
\Lambda_{\pm}
=
(\lambda+\epsilon)
\pm
(\lambda-\epsilon)|\cos\alpha|.
\end{equation}
Since the axis lengths scale as the square roots of the corresponding eigenvalues, their ratio is
\begin{equation}
\frac{L_{+}}{L_{-}}
=
\sqrt{\frac{\Lambda_{+}}{\Lambda_{-}}}
=
\sqrt{
\frac{
(\lambda+\epsilon)+(\lambda-\epsilon)|\cos\alpha|
}{
(\lambda+\epsilon)-(\lambda-\epsilon)|\cos\alpha|
}
}.
\end{equation}
The combined ellipse therefore becomes less elongated as $\alpha$ increases.

This result has a simple geometric connection to PCA, which finds the directions along which the projected data have the largest variance.
In two dimensions, the first PCA mode corresponds to the major axis of the ellipse, along which most of the variance is concentrated.
In a higher-dimensional data space, the ellipse generalizes to a hyper-ellipsoid, with the dominant PCA modes corresponding to its major principal directions.
We assume that these modes contain most of the foreground variance.
{When two datasets represented by hyper-ellipsoids with different orientations are added together, the resulting distribution becomes less elongated.}
The variance is therefore distributed over more principal directions, requiring more PCA modes to describe the correlated variation.
In the beam-rotation simulations, the two beam orientations correspond to two datasets.
Combining them can therefore make a fixed number of PCA modes less effective, leaving stronger correlated residuals.

\section{Common feature in adjacent pixels}
\label{app:Common feature in adjacent pixels}
\begin{figure}
    \centering
    \includegraphics[width=1\linewidth]{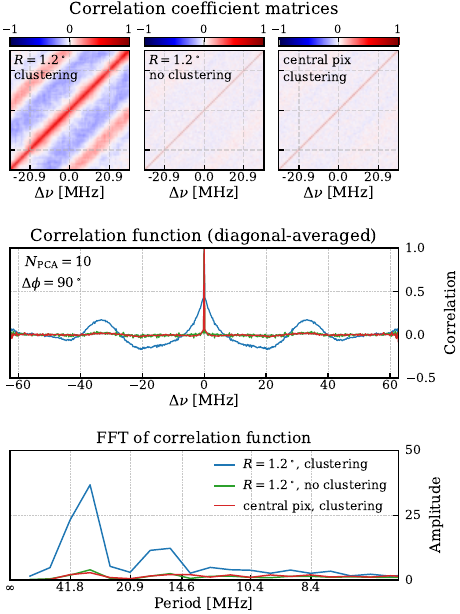}
    \caption{Upper panel: correlation matrices obtained from the simulation. 
    Middle panel: normalized two-point correlation functions obtained by averaging the corresponding correlation matrices along their diagonals. 
    Lower panel: the Fourier transforms of the corresponding correlation functions.
    }
    \label{fig:matrix_sp90}
\end{figure}

In this section, we show that the pixel-dependent frequency structure also appears in the simulations.
We consider different stacking ranges, with and without galaxy clustering.
In \autoref{fig:matrix_sp90}, we plot the correlation matrices of three different cases.
First, stacking radius is set to $1.2^\circ$ and galaxies are clustered.
Second, stacking radius is set to $1.2^\circ$ but galaxies are homogeneous.
The last, galaxies are clustered but only the central pixel is used to stack.
In the last two scenarios, the ripple pattern in the correlation matrix is significantly fainter than in the first one.
This implies that the oscillating structure is also pixel-dependent in simulation.

\bsp	
\label{lastpage}
\end{document}